\documentclass[final,twoside]{IEEEtran}
\usepackage{cite}
\ifCLASSINFOpdf
\usepackage[pdftex]{graphicx}
\DeclareGraphicsExtensions{.pdf,.jpeg,.png}
\else
\usepackage[dvips]{graphicx}
\DeclareGraphicsExtensions{.eps}
\fi

\usepackage[cmex10]{amsmath}
\usepackage{amsfonts}
\usepackage{amssymb}
\usepackage{amsthm}
\usepackage{pifont}
\usepackage{booktabs}
\usepackage{multirow}
\usepackage{color}
\usepackage[colorlinks=true,,linkcolor=black,citecolor=black,]{hyperref}
\newcommand{\cmark}{\ding{51}}%
\newtheorem{Theo}{Theorem}
\newtheorem{Coro}{Corollary}

\usepackage{algorithm}
\usepackage{algpseudocode}
\usepackage{algorithmicx}
\usepackage{rotating}
\usepackage{enumitem}
\usepackage{multirow}
\usepackage{array}
\usepackage{makecell}
\usepackage{color}

\usepackage{tikz,xcolor,hyperref}

\definecolor{lime}{HTML}{A6CE39}
\DeclareRobustCommand{\orcidicon}{%
	\begin{tikzpicture}
	\draw[lime, fill=lime] (0,0)
	circle [radius=0.16]
	node[white] {{\fontfamily{qag}\selectfont \tiny ID}};
	\draw[white, fill=white] (-0.0625,0.095)
	circle [radius=0.007];
	\end{tikzpicture}
	\hspace{-2mm}
}

\foreach \x in {A, ..., Z}{%
	\expandafter\xdef\csname orcid\x\endcsname{\noexpand\href{https://orcid.org/\csname orcidauthor\x\endcsname}{\noexpand\orcidicon}}
}

\ifCLASSOPTIONcompsoc
\usepackage[caption=false,font=normalsize,labelfont=sf,textfont=sf,farskip=0pt]{subfig}
\else
\usepackage[caption=false,font=footnotesize,farskip=0pt]{subfig}
\fi
\usepackage{fixltx2e}
\ifCLASSOPTIONcaptionsoff
\usepackage[nomarkers]{endfloat}
\let\MYoriglatexcaption\caption
\renewcommand{\caption}[2][\relax]{\MYoriglatexcaption[#2]{#2}}
\fi
\usepackage{url}
\begin{document}
	\title{SMIX($\lambda$): Enhancing Centralized Value \\Functions for Cooperative Multi-Agent Reinforcement Learning}

	\author{Xinghu~Yao$^*$,
		Chao~Wen$^*$, Yuhui~Wang
		and~Xiaoyang~Tan\\
		Nanjing University of Aeronautics and Astronautics, China\\
		\{xinghuyao, chaowen, y.wang, x.tan\}@nuaa.edu.cn
		\thanks{$^*$Equal contribution.}
		\thanks{Code is available at: https://github.com/chaovven/SMIX}
	}
	
	%



	\maketitle

\begin{abstract}
Learning a stable and generalizable centralized value function (CVF) is a crucial but challenging task in multi-agent reinforcement learning (MARL), as it has to deal with the issue that the joint action space increases exponentially with the number of agents in such scenarios. This paper proposes an approach, named SMIX(${\lambda}$), that uses an off-policy training to achieve this by avoiding the greedy assumption commonly made in CVF learning. As importance sampling for such off-policy training is both computationally costly and numerically unstable, we proposed to use the ${\lambda}$-return as a proxy to compute the TD error. With this new loss function objective, we adopt a modified QMIX network structure as the base to train our model. By further connecting it with the ${Q(\lambda)}$ approach from an unified expectation correction viewpoint, we show that the proposed SMIX(${\lambda}$) is equivalent to ${Q(\lambda)}$ and hence shares its convergence properties, while without being suffered from the aforementioned curse of dimensionality problem inherent in MARL. Experiments on the StarCraft Multi-Agent Challenge (SMAC) benchmark demonstrate that our approach not only outperforms several state-of-the-art MARL methods by a large margin, but also can be used as a general tool to improve the overall performance of other CTDE-type algorithms by enhancing their CVFs.
\end{abstract}




\section{Introduction}

\label{sec:introduction}

\IEEEPARstart{R}{ecently}, reinforcement learning (RL) has made great success in a variety of domains, from game playing \cite{mnih2015human,silver2017mastering} to complex continuous control tasks \cite{zhao2014mec,kiumarsi2017optimal,yang2018hierarchical}. However, many real-world problems are inherently multi-agent in nature, such as network packet routing \cite{ye2015multi}, automatic control \cite{van2016coordinated,liu2019cooperative}, social dilemmas \cite{yu2015emotional}, consensus in multi-agent systems \cite{jing2016consensus,bu2017data,zheng2017consensus} and multi-player video games \cite{jaderberg2019human}, which raises great challenges that are never encountered in single-agent settings.

In particular, the main challenges in multi-agent environments include the dimension of joint action space that grows exponentially with the number of agents \cite{foerster2018counterfactual,rashid2018qmix}, unstable environments caused by the interaction of individual agents \cite{laurent2011world,lowe2017multi}, and multi-agent credit assignment in cooperative scenarios with global rewards \cite{foerster2018counterfactual,rashid2018qmix}. These challenges make it troublesome for both fully centralized methods which consider all agents as a single meta agent and fully decentralized methods which individually train each agent by treating other agents as part of the environment.

Recently the paradigm of centralized training with decentralized execution (CTDE) has become popular for multi-agent reinforcement learning \cite{oliehoek2008optimal,kraemer2016multi,foerster2018counterfactual,rashid2018qmix} due to its conceptual simplicity and practical effectiveness. Its key idea is to learn a centralized value function (CVF) shared by all the agents during training, while each agent acts in a decentralized manner during the execution phase. The CVF works as a proxy to the environment for each agent, through which individual value/advantage functions for each agent can be conveniently learned by incorporating appropriate credit assignment mechanism.

Unfortunately, the central role played by the centralized value function in the CTDE approach seems to receive inadequate attention in current practice - it is commonly treated in the same way as in single-agent settings \cite{lowe2017multi,rashid2018qmix,foerster2018counterfactual,sunehag2018value}, leading to larger estimation error in multi-agent environments. Furthermore, to reduce the difficulty of decomposing the centralized value function to individual value functions, many algorithms impose extra structural assumptions onto the hypothesis space of the centralized value function during training. {For example, \textit{value decomposition networks} (VDN) \cite{sunehag2018value}, \textit{monotonic value function factorization} (QMIX) \cite{rashid2018qmix}, and \textit{factorizaton with transformation} (QTRAN) \cite{son2019qtran} assume that the optimal joint action is equivalent to the collection of each agent's optimal action.}

On the other hand, performing an accurate estimation of centralized value function in multi-agent environments is inherently difficult due to the following reasons: 1) the ``curse of dimensionality" \cite{bellman1957dynamic} of the joint action space results in the sparsity of experiences; 2) the partial observability in multi-agent environments become even more severe than in single-agent settings; 3) the dynamics of multi-agent environments are complex and hard to model, partially due to the complicated interactions among agents. In practice, these factors usually contribute to an unreliable and unstable centralized value function with high bias and variance.

{To tackle these difficulties, this work proposes a new sample efficient multi-agent reinforcement learning method, named SMIX($\lambda$), under the CTDE framework. We summarize our major contributions in the following threefold.
	
Firstly, we propose a general optimization framework for CTDE (Centralized Training with Decentralized Execution) in the context of fully cooperative MARL (Multi-Agent Reinforcement Learning) and analyze its theoretical properties. While many previous CTDE-type MARL methods are built on the centralized greedy behavior (CGB) assumption, which states that a set of decentralized greedy policies (one for each agent) are collectively optimal for the centralized greedy policy. We emphasize more on the importance of improving generalization capability of the learnt policies by incorporating useful inductive bias (such as the non-negative constraint for multi-agent coordination). Relaxing CGB \footnote{{By `relaxing CGB', we mean that in our framework the CGB assumption is not explicitly needed although it could be derived from some constraints such as the non-negative constraint. }} is not only practical, but gives us the flexibility to choose a wider range of methods to train the centralized value function (CVF) as well, considering that many existing methods (especially those Q-learning based, e.g., \cite{rashid2018qmix}) have to rely on this assumption for tractable optimization in high-dimensional action space.
	
Secondly, we propose a novel variant of off-policy SARSA($\lambda$) \cite{sutton2018reinforcement} algorithm for centralized value function estimation in the context of CTDE. The method is characterized by its capability to learn from multi-step lookahead data with improved sample efficiency, but does not suffer from the intensive computational cost originally involved in estimating the product of importance-sampling (IS) ratios in joint action space, hence being particularly suitable for multi-agent reinforcement learning. We further establish its convergence property by building its connection with the well-known $Q(\lambda)$ algorithm \cite{harutyunyan2016q}. It is shown that our IS-free off-policy SARSA($\lambda$) algorithm is general and is widely applicable to many other CTDE-type multi-agent algorithms such as {counterfactual multi-agent} (COMA) policy gradients \cite{foerster2018counterfactual}, VDN \cite{sunehag2018value} and QTRAN \cite{son2019qtran}, to improve their performance.
	
Last but not least, with the flexibility provided by the proposed optimization framework, we propose a novel cooperative MARL method named SMIX($\lambda$) to learn a set of decentralized policies, with each agent only knowing its own partial observation, action history and a joint reward shared by all agents. The method is based on the newly-developed IS-free off-policy SARSA($\lambda$) algorithm, incorporated within the enhanced QMIX architecture \cite{rashid2018qmix} (hence the name SMIX($\lambda$), meaning ``SARSA($\lambda$)-based MIXture networks''). It is shown that the proposed SMIX($\lambda$) approach is reliable, easy to implement and significantly outperforms several state-of-the-art CTDE methods including COMA, VDN, QMIX, and QTRAN, on the benchmark of the StarCraft Multi-Agent Challenge \cite{samvelyan2019starcraft}.

A preliminary version of this work appears in \cite{chaowen2020smix}. However, due to page limits, \cite{chaowen2020smix} fails to cover all important information about SMIX($\lambda$). This expanded version aims to help readers gain a more comprehensive understanding of SMIX($\lambda$). Specifically, we summarize oue major contrubution in the introduction section and the related work section (Section \ref{sec:relatedwork}) is added for intruducing related works. The Section \ref{sec:framework} is added for building a gerenal CTDE optimization framework.  Moreover, the detailed proofs of theorems and more derivation details are included in order to provide a more detailed description of the theoretical properties of QMIX, SMIX($\lambda$) and $Q(\lambda)$ \cite{harutyunyan2016q}. Besides, Figure \ref{architecture} illustrates that the estimation method of the CVF in SMIX($\lambda$) can be easily applied to other popular value-based and actor-critic-based CTDE methods. And more experimental results have been added to show the effectiveness and generality of this estimation method. Last but not least, more implementation details are presented in Algorithm \ref{alg:alg1} and Section \ref{sec:experiments}, which can improve the reproducibility of SMIX($\lambda$).

In what follows, we first introduce the related work in Section \ref{sec:relatedwork}, then after a brief discussion about CTDE-type methods, a general CTDE optimization framework is established in Section \ref{sec:framework}, the proposed SMIX($\lambda$) method and its theoretical analysis are respectively described in Section \ref{sec:methods} and Section \ref{sec:analysis}. Main experimental results and ablation studies are given in Section \ref{sec:experiments} and we conclude the paper in Section \ref{sec:conclusions}.}

\section{Related Work}
\label{sec:relatedwork}
Deep reinforcement learning (DRL) has made significant progress in recent years with the powerful representation capabilities of deep neural networks \cite{mnih2015human,silver2017mastering,schulman2017proximal}. However, challenges in multi-agent scenarios such as the unstable environments and curse of dimensionality make it hard to apply classic deep reinforcement learning methods to multi-agent environments \cite{foerster2018counterfactual,nguyen2020deep,hernandez2019survey}.

The \textit{centralized training with decentralized execution} (CTDE) paradigm provides a simple solution to the above issue by separating the agent learning and execution, under the greedy assumption that the optimal actions for individual agents lead to optimal joint action. It has gradually become the de facto standard in cooperative multi-agent scenarios due to its conceptual simplicity and practical effectiveness. {Representative methods include \textit{counterfactual multi-agent} (COMA) policy gradients \cite{foerster2018counterfactual}, \textit{value decomposition networks} (VDN) \cite{sunehag2018value}, \textit{monotonic value function factorization} (QMIX) \cite{rashid2018qmix}, and \textit{factorizaton with transformation} (QTRAN) \cite{son2019qtran} }-- COMA is an on-policy actor-critic method that uses a carefully designed counterfactual baseline to perform credit assignment, while VDN, QMIX, and QTRAN are typical value-based CTDE methods by learning individual agents through learning a centralized value function first.

{Our SMIX($\lambda$) belongs to the CTDE framework as well, but we focus more on how to improve the sample efficiency and how to perform an accurate estimation of the centralized value function.} Our key idea is to use off-policy training to achieve these goals while relaxing the greedy assumption in the learning stage. Although off-policy methods are known to improve the sample efficiency \cite{mnih2015human,schaul2015prioritized}, the popular importance sampling methods for off-policy training are problematic as these methods often involve calculating a product of a series of importance sampling ratios, which is not only computationally costly but has high variance \cite{munos2016safe} as well.


The estimation of the CVF plays a central role in the CTDE framework, as its bias and variance directly affect the performance of the whole system. Foerster et al. adopt a variant of TD($\lambda$) \cite{foerster2018counterfactual} to balance the bias and variance in CVF estimation, but they use an on-policy training method which could be sample inefficient. Precup et al. propose an importance-sampling-based TD($\lambda$) method in the single-agent setting and prove the convergence property with linear function approximation \cite{precup2001off}.

{ Under the fully decentralized framework, \cite{foerster2017stabilising} proposes two methods to stabilize the off-policy training process. For the first method, the authors proposed adding extra time tags onto every piece of information to be stored in the replay buffer. This allows to decay obsolete data. In the second method, each agent's value function is conditioned on a fingerprint that disambiguates the age of the data sampled from the replay memory. Both methods help alleviate the non-stationary problem, but we adopt an alternative mechanism for this rather than manipulating the replay memory directly, as discussed in Section~\ref{sec:non-stationarity}. }

It is worth mentioning that in practice, however, off-policy correction is not always needed in off-policy learning, especially when the behavior policy and target policy are close to each other. For example, Hernandez et al. find that it is possible to ignore off-policy correction over off-policy SARSA \cite{sutton2018reinforcement} and $Q(\sigma)$ \cite{de2018multi} without seeing an adverse effect on the overall performance \cite{hernandez2019understanding}. Fujimoto et al. show that the off-policy experiences generated during the interaction with the environment tend to be heavily correlated to the current policy, and their experimental results also reveal that the distribution of off-policy data during the training procedure is very close to that of the current policy \cite{fujimoto2019off}. Their analysis provides an intuitive explanation for why performance can be improved even without off-policy correction. {Unfortunately, a notable gap remains between the empirical success and the underlying theoretical support. In Theorem \ref{theo:matching} of Section \ref{sec:analysis}, we give a principled way to justify this `thumb of rule', showing that a computationally efficient experience replay method such as ours in the context of MARL is not only feasible but theoretically sound as well.}

{There are other ways to deal with multi-agent problems. For example, agents can exchange information with each other through a communication channel \cite{foerster2016learning,singh2019learning} or a shared network structure\cite{jiang2018learning,iqbal2019actor}. The opponent modeling methods aim to infer other agents' policies by interacting with the environment and observing other agents' policy \cite{foerster2018learning,letcher2019stable}. These methods are designed to establish explicit or implicit connections among agents, and we aim to coordinate each agent through a centralized value function. Thus, these approaches are complementary to ours. Besides, multi-agent self-play \cite{sukhbaatar2018intrinsic} and task decomposition \cite{sun2020reinforcement} have recently been shown to be useful in MARL .}

Finally, there have been several attempts on the StarCraft Multi-Agent Challenge (SMAC) \cite{samvelyan2019starcraft}, including \cite{foerster2017stabilising,foerster2018counterfactual,rashid2018qmix}. The results of \cite{rashid2018qmix} are the published state-of-the-art in value-based methods and \cite{foerster2018counterfactual} in actor-critic-based methods.

\section{The CTDE Optimization Framework}
\label{sec:framework}
\subsection{Problem Formulation}
The cooperative multi-agent task we considered can be described as a variant of \textit{decentralized partially observable Markov decision process} (Dec-POMDP) \cite{oliehoek2016concise}. Specifically, this task can be defined as a tuple: $\mathcal{G=\langle S,A,P},r,\mathcal{Z,O},N,\gamma\rangle$, where $s\in \mathcal{S}$ denotes the true state of the environment, $\mathcal{A}$ is the action set for each of $N$ agents, and $\gamma\in[0,1]$ is the discount factor. At each timestep, each agent $i\in\left\{1,2,\cdots,N\right\}$ chooses an action $a^i\in \mathcal{A}$, forming a joint action $\boldsymbol{a} = \left\{a^1,a^2,\cdots,a^N\right\}\in \mathcal{A}^N$. Then the environment gets into next state $s^{\prime}$ through a dynamic transition function $\mathcal{P}(s^{\prime}|s,\boldsymbol{a}): \mathcal{S}\times \mathcal{A}^N\times \mathcal{S} \mapsto [0,1]$. All agents share the same reward function $r(s,\boldsymbol{a}): \mathcal{S}\times \mathcal{A}^N\mapsto \mathbb{R}$.  We consider a partial observable scenario\footnote{In standard Dec-POMDP, the observation function $\mathcal{Z}(\mathbf{o}|\boldsymbol{a},s^{\prime})$ denotes the probability of the observing joint observation $\mathbf{o}$ given that joint action $\boldsymbol{a}$ was taken and led to state $s^{\prime}$ (cf., \cite{oliehoek2016concise}).} in which each agent draws partial observation $o\in \mathcal{O}$ from the observation function $\mathcal{Z}(s,i):\mathcal{S}\times N\mapsto \mathcal{O}$. Each agent $i$ also has an observation-action history $\tau^{i}\in \mathcal{T}\equiv(\mathcal{O}\times \mathcal{A})^{*}$, on which it conditions a stochastic policy. A stochastic policy is a mapping defined as $\pi(a|\tau):\mathcal{T}\times\mathcal{A}\mapsto[0,1]$.

In Dec-POMDP, the goal of a learning algorithm is to obtain a group of individual policies that can maximize the expected discounted returns $\mathbb{E}_{\boldsymbol{a}\in \boldsymbol{\pi}, s\in\mathcal{S}}\left[\sum_{t=0}^{\infty}\gamma^tr(s,\boldsymbol{a})\right]$. To simplify notation, we denote joint quantities over agents in bold. We also omit the index $i$ of each agent when there is no ambiguity in the following sections.

The \textit{centralized training with decentralized enecution (CTDE)} approach provides a solution to the Dec-POMDP problem by introducing a centralized structure to coordinate the decentralized policies. In the training phase of the CTDE paradigm, a centralized action-value function $Q([s,\boldsymbol{\tau}],\boldsymbol{a})$ (or simply expressed as $Q(\boldsymbol{\tau},\boldsymbol{a})$) is learned from the local observation history of all agents (denoted as $\boldsymbol{\tau} = \{\tau^1, \tau^2, \cdots, \tau^N\}$) and the global state (denoted as $s$), while during the execution phase, each agent's policy $\pi^i$ only relies on its own observation-action history $\tau^i$.

\subsection{{A General CTDE Optimization Framework}}
{In this paper, we consider a fully cooperative multi-agent scenario, in which the relationship between centralized value function and decentralized value functions satisfies particular coordinative constraints. Formally, we consider value-based CTDE as the following optimization problem.
\begin{equation}
\begin{aligned}
&\mathop{\text{maximize}}_{\boldsymbol{\pi} }\mathop{\mathbb{E}}_{s_0\sim \rho_0(s_0),a^1\sim \pi^1,\cdots, a^N\sim\pi^N}\left[{Q_{tot}^{\boldsymbol{\pi}}(s_0,\boldsymbol{a})}\right] \\
&\text{subject to }  Q_{tot}(\boldsymbol{\tau},\boldsymbol{a}) = f(Q^1(\tau^1,a^1),\cdots,Q^N(\tau^N,a^N)),
\end{aligned}\label{general_optimization problem}
\end{equation}
where $f$ is a \textit{state-dependent continuous} function, $\rho_0:\mathcal{S}\rightarrow [0,1]$ is the distribution of the initial state $s_0$. The value function of each agent are combined to give the centralized value function $Q_{tot}$, through a \textit{state-dependent continuous} function $f$ implemented as a neural network.


Designing a proper optimization constraint in (\ref{general_optimization problem}) is crucial because it directly affects the generalization ability and optimization cost of the algorithm. In this paper, we assume the relationship between centralized value function and decentralized functions satisfies the following assumption.

{\textbf{Assumption 1.} In fully coorperative multi-agent scenarios, the coordination among agents can be constrained through $\frac{\partial Q_{t o t}}{\partial Q^i} \geq 0, i \in \{1, \cdots, N\}$.}

This non-negative constraint (positive weight) first appears in VDN \cite{sunehag2018value} and is generalized in QMIX \cite{rashid2018qmix} as a way to ensure tractable optimization in MARL. In this paper, however, we take it as a prior to capture the coordination information required to solve cooperative tasks. This effectively strikes a balance between coordinative expressivity and learning difficulty. As a result, each agent has a chance to influence the team reward positively instead of canceling out with each other in fully cooperative scenarios.


The following theorem guarantees that if assumption 1 is satisfied, then the optimal policy of each agent is conditioned on the optimal joint actions of other agents through the centralized observable critic $Q_{tot}^{\boldsymbol{\pi}}$.
This helps to address the non-stationary issue of MARL, as discussed later.
\begin{Theo}
	\label{theo1}
	In optimation problem (\ref{general_optimization problem}) with assumption 1 satisfied, if the joint action-value function $Q_{tot}^{\boldsymbol{\pi}}$ is already obtained, then we have:
	\begin{equation}
	\mathop{\operatorname{argmax}}_{a^i}Q^i(\tau^i,a^i) = \mathop{\operatorname{argmax}}_{a^i}\max_{a^1,\cdots,a^{i-1},a^{i+1},\cdots,a^N}Q^{\boldsymbol{\pi}}_{tot}(\boldsymbol{\tau},\boldsymbol{a})
	\end{equation}
\end{Theo}
\begin{IEEEproof}
	We denote $a^{1*} = \operatorname{argmax}_{a^1}Q^1(\tau^1,a^1)$. Since $\frac{\partial Q_{tot}^{\boldsymbol{\pi}}} { \partial Q^1} \geq 0$, we have
	\begin{equation*}
	\begin{aligned} &\max_{a^1,\cdots,a^{i-1},a^{i+1},\cdots,a^{N}}Q_{tot}^{\boldsymbol{\pi}}(\boldsymbol{\tau},a^1,a^{i-1},a^i,a^{i+1},\cdots,a^N)\\
	&=\max_{a^2,\cdots,a^{i-1},a^{i+1},\cdots,a^N}Q_{tot}^{\boldsymbol{\pi}}(\boldsymbol{\tau},a^{1*},a^{i-1},a^i,a^{i+1},\cdots,a^N).
	\end{aligned}
	\end{equation*}
	Similarly, we denote $a^{j*}=\operatorname{argmax}_{a^j}Q^j(\tau^j,a^j)$ for $j\in\{2,3,i-1,i+1,N\}.$ Then, due to $\frac{\partial Q_{tot}^{\boldsymbol{\pi}}} { \partial Q^j} \geq 0$, we have
	\begin{small}
		\begin{equation*}
		\begin{aligned}
		&\max_{a^2,\cdots,a^{i-1},a^{i+1},\cdots,a^{N}}Q_{tot}^{\boldsymbol{\pi}}(\boldsymbol{\tau},a^{1*},\cdots,a^{i-1},a^i,a^{i+1},\cdots,a^N)\\
		&=\max_{a^3,\cdots,a^{i-1},a^{i+1},\cdots,a^N}Q_{tot}^{\boldsymbol{\pi}}(\boldsymbol{\tau},a^{1*},a^{2*},a^{i-1},a^i,a^{i+1},\cdots,a^N)\\
		&=\cdots\\
		&=Q_{tot}^{\boldsymbol{\pi}}(\boldsymbol{\tau},a^{1*},a^{2*},a^{(i-1)*},a^i,a^{(i+1)*},\cdots,a^{N*}).
		\end{aligned}
		\end{equation*}
	\end{small}
	Then, given $\frac{\partial Q_{tot}^{\boldsymbol{\pi}}} { \partial Q^i} \geq 0$, we have
	\begin{equation*}
	\begin{aligned}
	&\operatorname{argmax}_{a^i}Q_{tot}^{\boldsymbol{\pi}}(\boldsymbol{\tau},a^{1*},a^{2*},a^{(i-1)*},a^i,a^{(i+1)*},\cdots,a^{N*})\\
	&=\operatorname{argmax}_{a^i}Q^i(\tau^i,a^i)=a^{i*}.
	\end{aligned}
	\end{equation*}
\end{IEEEproof}


{In an idealized setting where each agent observes the full state (each agent's partial observation $\tau^{i}$ is equivalent to the global state $s$), the optimal joint action-value function $Q^*$ is a solution of a \textit{multi-agent Markov decision process} (MMDP) which is itself equivalent to a standard MDP with $\mathcal{A}^N$ as the action space \cite{boutilier1996planning}. In such cases, the joint action $\boldsymbol{a_*}=\mathop{\operatorname{argmax}}_{\boldsymbol{a}}Q^*(\boldsymbol{\tau},\boldsymbol{a})$ is an \textit{Nash equilibrium} from the view of game theory \cite{hu2003nash}. Specifically, a Nash equilibrium is achieved among a group of agent if the following expression holds for all $i\in\{1,\cdots,N\}$:
\begin{equation}
{Q}^*(\boldsymbol{\tau},\boldsymbol{a_*}) = {Q}^*(\boldsymbol{\tau},a^{i}_*,\boldsymbol{a}^{-i}_*)\geq {Q}^*(\boldsymbol{\tau},a^{i},\boldsymbol{a}^{-i}_*), \forall a^i \in \mathcal{A},\label{eq:nash}
\end{equation}
where each agent acts with the \textit{best response} $a_*^i$ to others.

We adopt a compact notation for the joint action of all agents except $i$ as $\boldsymbol{a}_{*}^{-i}\triangleq\left[a_*^1,\cdots,a_*^{i-1},a_*^{i+1},\cdots,a_*^{N}\right].$ The following theorem shows that the \textit{Nash equilibrium} $\boldsymbol{a}_*$ can be obtained when each agent $i$ acts greedy according to $Q^{i}({\tau^i},a^i)$ if the optimal centralized value function $Q^*(\boldsymbol{\tau},\boldsymbol{a})$ satisfies assumption 1.
\begin{Coro}
	If the optimal joint action-value function $Q^*$ satisfies assumption 1, then the optimal joint policy $\boldsymbol{a}_* = \{a^{1*},a^{2*},\cdots,a^{N*}\}$, where $a^{i*}=\mathop{\operatorname{argmax}}_{a^i}Q^i(\tau^i,a^i)$ and $Q^i(\tau^i,a^i)$ is the decentralized value function obtained by solving (\ref{general_optimization problem}).
\end{Coro}
\begin{IEEEproof}
	According to theorem \ref{theo1}, we have
	$a^{i*}=\mathop{\operatorname{argmax}}_{a^i}Q_{tot}^{\boldsymbol{\pi}^*}(\boldsymbol{\tau},a^{i},a^{-i*}).$
	Thus, we have:
	\begin{equation*}
	Q_{tot}^{\boldsymbol{\pi}^*}(\boldsymbol{\tau},a^{i*},a^{-i*})\geq Q_{tot}^{\boldsymbol{\pi}^*}(\boldsymbol{\tau},a^{i},a^{-i*}), \forall a^i\in\mathcal{A}.
	\end{equation*}
	The above expression holds for all $i\in\{1,2,\cdots,N\}$.
\end{IEEEproof}}


\subsection{{CTDE Under Centralized Greedy Behavior Assumption}}
\label{sec:CGB}
To facilitate the freedom of each agent to make decision based on its local observation without consulting the centralized value function, the following centralized greedy behavior (CGB) assumption is usually adopted:
\begin{equation}
\underset{\boldsymbol{a}}{\operatorname{argmax}}\ Q_{tot}(\boldsymbol{\tau}, \boldsymbol{a})=\left(\begin{array}{c}{\mathop{\operatorname{argmax}}\limits_{a^{1}} Q^{1}\left(\tau^{1}, a^{1}\right)} \\ {\vdots} \\ {\mathop{\operatorname{argmax}}\limits_{a^{N}} Q^{N}\left(\tau^{N}, a^{N}\right)}\end{array}\right)
.\label{eq:CGB}
\end{equation}
This assumption establishes a structural constraint between the centralized value function and the decentralized value functions, which can be thought of as a simplified credit assignment mechanism during the execution phase.

Many methods aim to impose a structure constraint between centralized value function and decentralized value functions to ensure the CGB assumption is satisfied. Specifically, VDN \cite{sunehag2018value} uses the following additive combination:
\begin{equation}
Q_{tot}(\boldsymbol{\tau},\boldsymbol{a}; \boldsymbol{\theta}) = \sum_{i=1}^{N}\alpha_iQ^i(\tau^i,a^i;\theta^i), \alpha_i\geq 0,\label{eq:vdnlinear}
\end{equation}
where $\boldsymbol{\theta}$ is the collection of the parameter vector $
\theta^i$ for each agent's action-value function.
In VDN \cite{sunehag2018value}, all the combination coefficients $\alpha_i,i=1, 2, \cdots, N$ are set to $1$. QMIX \cite{rashid2018qmix} extends this additive value factorization to a more general case by directly enforcing
$\frac{\partial Q_{t o t}}{\partial Q^i} \geq 0, i \in \{1, \cdots, N\}$ via a non-negative mixing network $f$.  With this, we can easily obtain the following theorem.
\begin{Theo}
	{For QMIX, we have}
	\begin{equation}\label{eq_greedy_gloableqlocal}
	\begin{aligned}
	&\max_{\boldsymbol{a}} Q_{tot}(\boldsymbol{\tau},\boldsymbol{a}) = \\&f\left(\boldsymbol{\tau},
	\mathop{\operatorname{argmax}}\limits_{a^1} Q^1(\tau^1,a^1), \cdots, \mathop{\operatorname{argmax}}\limits_{a^N} Q^N(\tau^N,a^N)\right).
	\end{aligned}
	\end{equation}\label{theo:theorem2}
\end{Theo}
%
\begin{IEEEproof}
	
	{Due to $Q_{tot}(\boldsymbol{\tau},\boldsymbol{a}) = f\left(  Q^1(\tau^1, a^1), \ldots, Q^N(\tau^N, a^N) \right)$,
		where $f$ is a state-dependent continuous monotonic function, and $\boldsymbol{a}=(a_1, \ldots, a_N)$.} Thus, we have
	\begin{equation*}
	\begin{aligned}
	Q_{tot}&\left(\tau,
	\operatorname{argmax}_{a^1} Q^1(\tau^1,a^1), \cdots, \operatorname{argmax}_{a^N} Q^N(\tau^N,a^N)\right)
	\\&=
	f\left( \max_{a^1}Q^1(\tau^1, a^1), \ldots, \max_{a^n} Q^N(\tau^N, a^N) \right).
	\end{aligned}
	\end{equation*}
	Since $\frac {\partial f} { \partial Q^i } \geq 0 $,
	{given } $(\bar a^2, \ldots, \bar a^n)$, {we have }
	\begin{equation*}
	\begin{aligned}
	f&\left(  Q^1(\tau^1, a^1), Q^2(\tau^2, \bar a^2), \ldots, Q^N(\tau^N, \bar a^N) \right)
	\\&\leq
	f\left( \max_{a^1}  Q^1(\tau^1, a^1),   Q^2(\tau^2, \bar a^2), \ldots, Q^N(\tau^N, \bar a^N) \right)
	\end{aligned}
	\end{equation*}
	for any $a^1$.
	Similarly, given $(\bar a^1, \ldots, \bar a^{k-1}, \bar a^{k+1}, \ldots, \bar a^N)$, we have
	\begin{equation*}
	\begin{aligned}
	f&\left(  Q_1(\tau^1, \bar a^1), \ldots, Q^k(\tau^k,  a^k), \ldots, Q^N(\tau^N, \bar a^N) \right)
	\\&\leq
	f\left(   Q^1(\tau^1, \bar a^1), \ldots, \max_{a^k}  Q^k(\tau^k, a^k) , \ldots, Q^N(\tau^N, \bar  a^N) \right)
	\end{aligned}
	\end{equation*}
	for any $a^k$.
	Finally, for any $( a^1, \ldots,  a^N)$, we have
	\begin{equation*}
	\begin{aligned}
	f&\left(  Q^1(\tau^1,  a^1), \ldots, Q^N(\tau^N,  a^N) \right)
	\\\leq &f\left( \max_{a^1} Q_1(\tau^1,  a^1), \ldots, Q^N(\tau^N,  a^N) \right)
	\\
	\leq & f\left( \max_{a^1} Q^1(\tau^1,  a^1), \ldots, \max_{a^N} Q^N(\tau^N,  a^N) \right).
	\end{aligned}
	\end{equation*}
	Therefore, we obtain
	\begin{equation*}
	\begin{aligned}
	\max_{a^1,\ldots, a^N} &f\left(  Q^1(\tau^1, a^1), \ldots, Q^N(\tau^N, a^N) \right) \\&=
	f\left( \max_{a^1} Q^1(\tau^1,  a^1),  \ldots, \max_{a^N} Q^N(\tau^N,  a^N) \right),
	\end{aligned}
	\end{equation*}
	which is the specific form of \eqref{eq_greedy_gloableqlocal}. \qedhere
\end{IEEEproof}

{ The above theorem shows that with the help of a non-negative mixing network $f$, performing Q-learning is tractable for $Q_{tot}(\boldsymbol{\tau},\boldsymbol{a})$ even in high dimensional joint action space. So does VDN. However, this non-negative constraint is only a sufficient condition of the CGB assumption, hence restricting the algorithm's representational complexity. To achieve the best generalization, QTRAN \cite{son2019qtran} allows to search the best hypothesis in a space equivalent to the one specified by the CGB condition. In other words, QTRAN works in a larger hypothesis space than both VDN and QMIX. However, optimizing in a larger hypothesis space requires more optimization efforts.} Although a coordinate-decent-type method is proposed in QTRAN to address this issue, the method's scalability and the range of practical use can be limited. See more discussions on this in the experimental section.}

	

\subsection{{Dealing with Non-Stationarity}} \label{sec:non-stationarity}
{ 

The non-stationary problem is a key issue in the MARL setting \cite{papoudakis2019dealing}. An environment that contains multiple agents, seen from the angle of an individual agent, is constantly changing with the changes of any other agent. Through this lens, the CTDE (Centralized Training with Decentralized Execution) scheme provides a simple solution to this issue by introducing a fully observable critic (i.e., the centralized value function $Q_{tot}$). The fully observable critic accesses to the observation and actions of all agents, and consequently the environment becomes stationary even though the policy of other agents changes. The fully observable critic also helps to simplify the optimization problem and enables the algorithm to scale better to more complex scenarios.
	
Alternatively, the CTDE strategy can be thought of as a way to condition each agent's policy on other agents' joint policy through the centralized observable critic (cf. Theorem 1). But this idea can also be implemented in a more straightforward way under fully decentralized settings without the need of the centralized observable critic, as in \textit{Hyper Q-learning}\cite{tesauro2004extending}, \textit{Distributed Q-learning}\cite{lauer2000algorithm} and \textit{multi-agent fingerprints}\cite{foerster2017stabilising}. Specifically, in \textit{Hyper Q learning}, each agent learns a Q function over all possible opponent strategies. Similarly, the \textit{Distributed Q-learning}\cite{lauer2000algorithm} considers a decentralized cooperative multi-agent problem in fully observable environments and the joint action is available for all agents in the training time. The \textit{Distributed Q-learning} algorithm updates the Q-values only when there is a guaranteed improvement. In \textit{multi-agent fingerprints}, each agent's value function conditions on a fingerprint that disambiguates the age of the data sampled from the replay memory, which helps to stabilize the training process in fully decentralized settings.
	
}

\begin{figure*}[htb!]
	\centering
	\includegraphics[width=\textwidth]{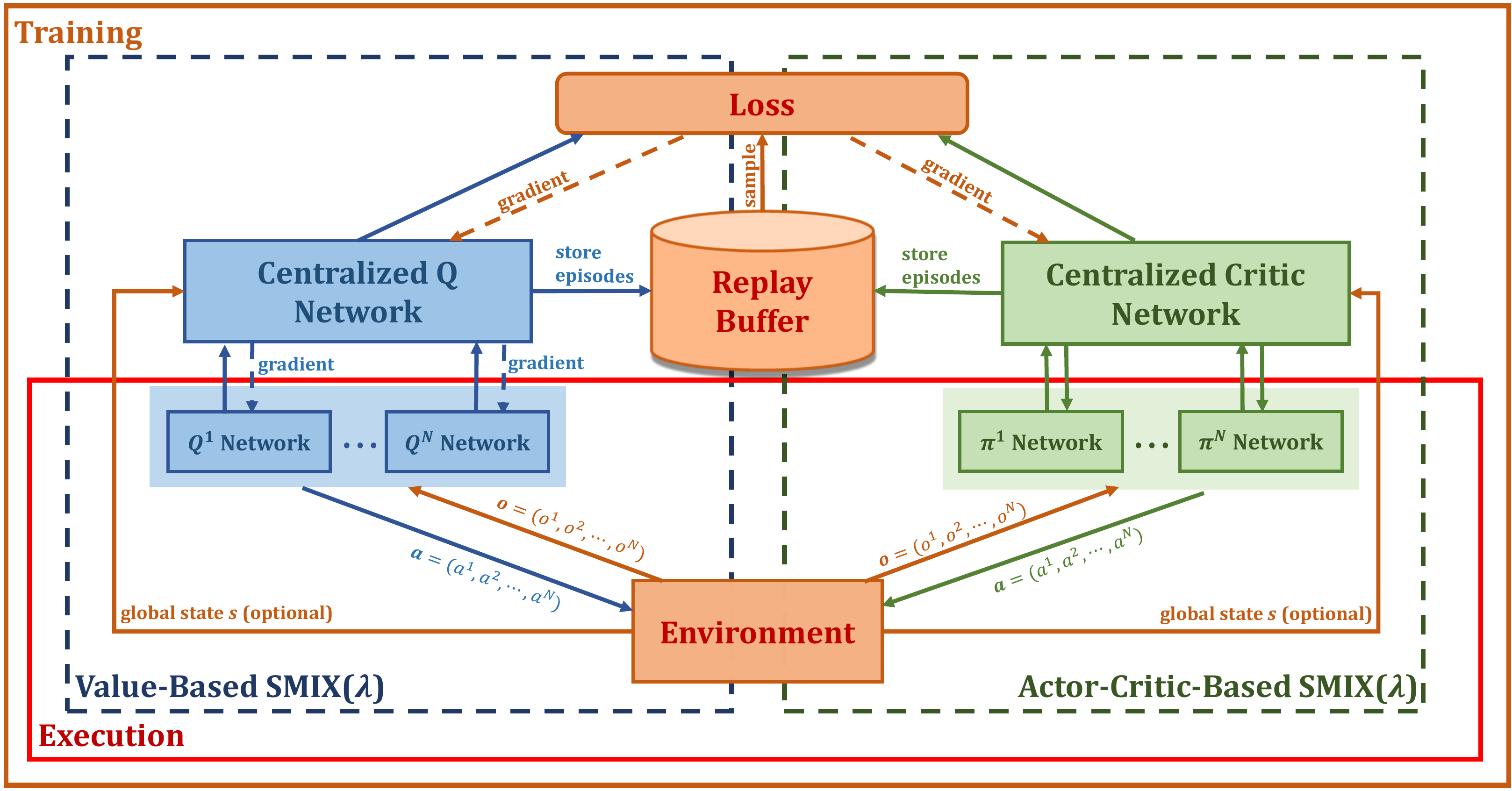}
	\caption{{Applying IS-free SARSA($\lambda$) to the centralized value function estimation for value-based and actor-critic-based methods.
 	(The left and right dashed boxes show the value-based SMIX($\lambda$) and actor-critic based SMIX($\lambda$) algorithms, and the two solid boxes represent the modules involved in the centralized training and decentralized execution respectively. Each agent's $Q$ network (or $\pi$ network) only has access to its own observation, and the centralized $Q$ network (or critic network) aggregates all agents' observation information.)   }}\label{architecture}
\end{figure*}

\section{Methods}
\label{sec:methods}

{In this section, we give the details of the proposed SMIX($\lambda$) method, which is a SARSA($\lambda$) \cite{sutton2014new} style off-policy method that aims at learning a flexible and generalizable centralized value function within the CTDE framework.

\subsection{Motivation}
One of the most popular methods to estimate $Q_{tot}^{\boldsymbol{\pi}}$ in (\ref{general_optimization problem}) is based on the Q-learning, as in VDN, QMIX, and QTRAN. However, to make this computationally tractable in a joint action space, as discussed in \ref{sec:CGB}, the CGB assumption has to be made. This potentially restricts the range of possible learning algorithms for solving (\ref{general_optimization problem}).


Alternatively, one can use an Expected-SARSA-based method. The Expected SARSA estimates its TD (temporal difference) target reinforcement signal with an expectation value of the next state-action pairs in an on-policy way\cite{sutton2018reinforcement}. In other words, it does not have to perform a greedy search over the joint action space and hence does not need the CGB assumption either. This allows us to decouple the learning algorithm and the CGB assumption. Through an iteration process over $Q_{tot}^{\boldsymbol{\pi}}$, the optimal joint action-value function can be obtained\cite{van2009theoretical}. Besides, it is well-known that the Q-learning algorithm can be viewed as a particular case of Expected SARSA in which the expectation over actions is replaced with a deterministically greedy one\cite{sutton2018reinforcement,de2018multi}.

In what follows, we will propose a new off-policy value function estimation method based on this idea and apply it to centralized value function estimation within the CTDE framework.}

\subsection{{Importance-Sampling-Free Off-policy SARSA($\lambda$)}}
\label{subsec:offpolicy}
Denoting the behavior policy as $\boldsymbol{\mu}$ and the target policy as $\boldsymbol{\pi}$, a general off-policy strategy to evaluate the $Q$ value function {(centralized value function $Q_{tot}$ in multi-agent setting)} for $\boldsymbol{\pi}$ using data $\boldsymbol{\tau}$ generated by following $\boldsymbol{\mu}$ can be expressed as follows \cite{munos2016safe},
\begin{equation}
Q(\boldsymbol{\tau}, \boldsymbol{a}) \leftarrow Q(\boldsymbol{\tau}, \boldsymbol{a})+\mathbb{E}_{\boldsymbol{\mu}}{\left[\sum_{t \geq 0} \gamma^{t}\left(\prod_{i=1}^{t} \rho_i\right)\delta_t^{\boldsymbol{\pi}}\right]},\label{eq:offpolicyoperator}
\end{equation}
where each $\rho_i$ is a non-negative coefficient and satisfies $\prod_{i=1}^{t} \rho_i=1$ when $t=0$. The error term $\delta_t^{\boldsymbol{\pi}}$ is generally written as the following expected TD-error,
\begin{equation}
\delta_t^{\boldsymbol{\pi}}={\underbrace{r_{t+1}+\gamma \mathbb{E}_{\boldsymbol{\pi}} Q\left(\boldsymbol{\tau}_{t+1}, \cdot\right)}_{{\text{1-step TD-target}}}-Q\left(\boldsymbol{\tau}_{t}, \boldsymbol{a}_{t}\right)},\label{eq:delta}
\end{equation}
where $\mathbb{E}_{\boldsymbol{\pi}}Q(\boldsymbol{\tau},\cdot)=\sum_{\boldsymbol{a}}\boldsymbol{\pi}(\boldsymbol{a}|\boldsymbol{\tau})Q(\boldsymbol{\tau},\boldsymbol{a})$.\footnote{The policy evaluation strategy of many popular methods can be expressed as (\ref{eq:offpolicyoperator}), including SARSA($\lambda$) \cite{sutton2014new}, off-policy importance sampling methods \cite{precup2001off}, off-policy $Q(\lambda)$ method \cite{harutyunyan2016q}, tree-backup method, TB($\lambda$) \cite{precup2000eligibility} and Retrace($\lambda$) \cite{munos2016safe}. These methods differ in the definition of the coefficient $\rho_i$ and error term $\delta_t^{\boldsymbol{\pi}}$ \cite{harutyunyan2016q,munos2016safe}.} In particular, for the importance sampling (IS) method, each $\rho_i$ in (\ref{eq:offpolicyoperator}) is defined as the relative probability of their trajectories occurring under the target policy $\boldsymbol{\pi}$ and behavior policy $\boldsymbol{\mu}$, also called importance sampling ratio, i.e., $\rho_i=\frac{\boldsymbol{\pi}(\boldsymbol{a}_i|\boldsymbol{\tau}_i)}{\boldsymbol{\mu}(\boldsymbol{a}_i|\boldsymbol{\tau}_i)}$.

Despite its theoretical soundness, the importance sampling (IS) method faces great challenges under the setting of multi-agent environments: 1) it suffers from large variance due to the product of the ratio \cite{liu2018breaking}, and 2) the ``curse of dimensionality" issue of the joint action space makes it impractical to calculate the $\boldsymbol{\pi}(\boldsymbol{a}_i|\boldsymbol{\tau}_i)$ even for a single timestep $i$, when the number of agents is large. Previously, Liu et al. proposed a method that effectively addresses the first issue by avoiding calculating the product over the trajectories \cite{liu2018breaking}, but how to solve the second one remains open.

The above analysis highlights the need for exploring alternative approaches that can perform off-policy learning without importance sampling in multi-agent settings. To achieve the above goal, the key idea of SMIX($\lambda$) is to further simplify the coefficient $\rho_i$ in (\ref{eq:offpolicyoperator}), so as to reduce the variance of the importance sampling estimator and to potentially bypass the curse of dimensionality involved in calculating $\boldsymbol{\pi}(\cdot|\boldsymbol{\tau})$.

Specifically, we relax each coefficient $\rho_i = 1.0$ in (\ref{eq:offpolicyoperator}) use the $\lambda$-return \cite{sutton2018reinforcement} as the TD target estimator, which is defined as follows:
\begin{equation}
{G}_{t}^{\lambda} = (1-\lambda)\sum_{n=1}^{\infty}\lambda^{n-1}{G}_{t}^{(n)},\label{eq:offpolicyreturn}
\end{equation}
where $	{G}_t^{(n)} = r_{t+1} + \gamma r_{t+2} +\cdots + \gamma^n\mathbb{E}_{\boldsymbol{\pi}} Q\left(\boldsymbol{\tau}_{t+n}, \boldsymbol{a}_{t+n}; \theta^{-}\right)$ is the $n$-step return and $\theta^{-}$ are the parameters of the target network.

{Replacing 1-step TD-target in (\ref{eq:delta}) with $G_{t}^{\lambda}$, and setting $\rho_i = 1.0$ for all $i$ in (\ref{eq:offpolicyoperator}), we have (the update step-size $\alpha$ is omitted for simplification)},
\begin{equation}
Q(\boldsymbol{\tau}, \boldsymbol{a}) \leftarrow Q(\boldsymbol{\tau}, \boldsymbol{a})+\mathbb{E}_{\boldsymbol{\mu}}{\left[\sum_{t \geq 0} \gamma^{t}({G}_{t}^{\lambda}-Q(\boldsymbol{\tau}_t,\boldsymbol{a}_t))\right]}.\label{eq:SMIXlambdaupdate}
\end{equation}

{In this paper, we call the method using (\ref{eq:SMIXlambdaupdate}) as the TD-target for off-policy value function estimation as ``IS-free off-policy SARSA($\lambda$)". Next, we use this method to estimate the centralization value function in multi-agent reinforcement learning and analyze its theoretical properties of this method in section \ref{sec:analysis}.}

\subsection{The SMIX($\lambda$) Algorithm}
\label{subsec:smixmethod}

SMIX($\lambda$) is trained end-to-end and the loss function for the centralized value function $Q^{\boldsymbol{\pi}}_{tot}$ has the following form:
\begin{equation}
\mathcal{L}_t(\theta) = \sum_{i=1}^{N_b}\left[(y_{i}^{tot}-Q_{tot}^{\boldsymbol{\pi}}(\boldsymbol{\tau},\boldsymbol{a}; \theta))^2\right],\label{eq:loss}
\end{equation}
where $y_i^{tot} = {G}_t^{\lambda}$ is defined in (\ref{eq:offpolicyreturn}) and is estimated through experience replaying, $N_b$ is the batch size.

In implementation, SMIX($\lambda$) use an experience replay \cite{mnih2015human} to store the most recent off-policy data. {The experience replay usually stores a queue of experiences (tuples of [observation, action, reward, successor observation]). In SMIX($\lambda$), however, each tuple in the experience replay stores one complete trajectory $(s_0,\boldsymbol{\tau}_0,\boldsymbol{a}_0,r_1,\cdots,s_{T-1},\boldsymbol{\tau}_{T-1},\boldsymbol{a}_{T-1}, r_T, s_{T})$ so as to evaluate the $\lambda$-return target.}

The QMIX \cite{rashid2018qmix} structure is adopted as the basic deep network architecture for the proposed SMIX($\lambda$). Each agent $i$ has its own decentralized $Q^i(\tau^i,a^i)$ network composed of GRU \cite{GRU} modules. Then all the individual $Q^i$ values are passed into a mixing network to calculate the joint action-value $Q_{\text{tot}}^{\boldsymbol{\pi}}$. {The weight of the mixing network is generated by hypernetworks\cite{ha2017hypernetworks} using the global state $s$.} All the neural networks are trained end-to-end and the centralized value function $Q_{\text{tot}}^{\boldsymbol{\pi}}$ is updated by minimizing the loss (\ref{eq:loss}).

The general training procedure for SMIX($\lambda$) is provided in Algorithm \ref{alg1}. {Firstly, the replay buffer is filled with the trajectories and the oldest data is replaced when the buffer is full. Secondly, we sample a batch of episodes uniformly from the replay buffer to calculate the $\lambda$-return TD target. Then, the parameters of behavior network $\theta$ are updated by minimizing the loss function. Finally, we replace the target network's parameters  $\theta^{-}$ with $\theta$ periodically and iterate the above process.}

It is worth noting that our method of training a centralized value function is a general method and can be easily applied to other CTDE methods, include value-based methods (such as VDN \cite{sunehag2018value}), actor-critic-based methods (e.g. COMA \cite{foerster2018counterfactual}), and even fully decentralized methods (e.g. {independent Q-learning (IQL) \cite{tan1993multi})}. Figure \ref{architecture} gives the overall architecture of a generalized version of SMIX($\lambda$).

\begin{algorithm*}[htb!]
	\caption{Training Procedure for SMIX($\lambda$)}\label{alg1}
	\begin{algorithmic}[1]
		\State Initialize the behavior network with parameters $\theta$, the target network with parameters $\theta^-$, empty replay buffer $\mathcal{D}$ to capacity $N_\mathcal{D}$, training batch size $N_b$
		
		\For{each training episode}
		\For{each episode}
		\For{$t=1$ to $T-1$}
		\State Obtain the partial observation $\mathbf{o}_t = \{o^1_t,\cdots,o^N_t\}$ for all agents and global state $s_t$
		\State Select action $a^i_t$
		according to $\epsilon$-greedy policy w.r.t agent $i$'s decentralized value function $Q^i$ for $i=1,\cdots,N$
		\State Execute joint action
		$\boldsymbol{a}_t=\{a^1,a^2,\cdots,a^N\}$ in the environment
		\State Obtain the global reward $r_{t+1}$, the next partial observation ${o^i_{t+1}}$ for each agent $i$ and next global state $s_{t+1}$
		\EndFor
		\State Store the episode in $\mathcal{D}$, replacing the oldest episode if $|\mathcal{D}| \geq N_\mathcal{D}$
		\EndFor
		\State Sample a batch of $N_b$ episodes $\sim \text{Uniform}(\mathcal{D})$
		\State Calculate  $\lambda$-return targets $y_i^{tot}$ according to (\ref{eq:offpolicyreturn}) using $\theta^{-}$ for each timestep
		\State Update $\theta$ by minimizing $\sum_{t=1}^{T-1}\sum_{i=1}^{N_b}\left[(y_{i}^{tot}-Q_{tot}^{\boldsymbol{\pi}}(\boldsymbol{\tau},\boldsymbol{a}; \theta))^2\right]$
		\State Replace target parameters $\theta^- \leftarrow \theta$ every $C$ episodes
		
		\EndFor
	\end{algorithmic}\label{alg:alg1}
\end{algorithm*}
\subsection{Representational Complexity}
\label{sec:representational}
\begin{figure}[htb!]
	\centering
	\includegraphics[width=0.45\textwidth]{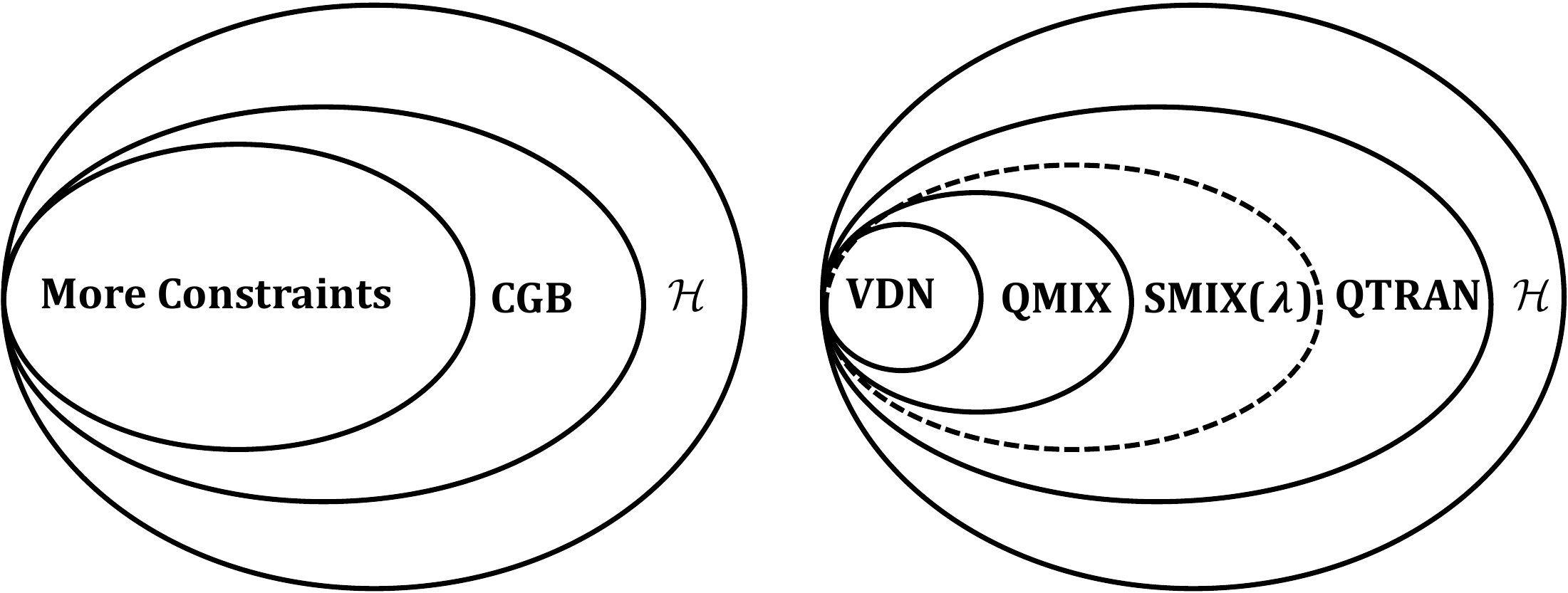}
	\text{~~~~~~~~~~~~~~~~}(a) \hfill (b) \text{~~~~~~~~~~~~~~}
	\caption{(a) The size of hypothesis space corresponding to different constraints. (b) {The relationship of representational complexity for several different algorithms.}}\label{CGB}
\end{figure}

The hypothesis space (or hypothesis set) $\mathcal{H}$ is a space of all possible hypotheses for mapping inputs to outputs that can be searched \cite{shalev2014understanding,russell2016artificial}. To learn a stable and generalizable CVF, choosing a suitable hypothesis space is essential, which is related not only to the characteristic of the problem domain but also to how the learned system is deployed. In particular, in multi-agent systems, all agents' joint action space increases exponentially with the increase of the number of agents, implying that the hypothesis space of CVF should be large enough to account for such complexity. {However, to reduce optimization costs and enable decentralized execution, it is often necessary to impose some constraints on centrally valued functions.} Figure \ref{CGB}(a) shows the relationship of hypothesis spaces under some different constraints.

{The centralized value function $Q_{tot}$ obtained by solving optimization problem (\ref{general_optimization problem}) can represent any function that fits assumption 1. Our SMIX($\lambda$) also uses the non-negative constraint during training. This makes QMIX\cite{rashid2018qmix} and SMIX($\lambda$) share the same representational complexity of centralized value function. However, SMIX($\lambda$) is more flexible than QMIX due to the decoupling of updating rule and CGB assumption during training. QTRAN \cite{son2019qtran} directly optimizes the joint action-value function, which gives this method a stronger representational complexity than QMIX and SMIX($\lambda$). Figure \ref{CGB}(b) shows the relationship of representational complexity for several different algorithms.}

{We use an $m$-step cooperative matrix game \cite{mahajan2019maven} for two agents to illustrate the effects of representational complexity of QMIX, SMIX($\lambda$) and QTRAN. Similar matrix games are usually used to study the algorithm's representational complexity \cite{rashid2018qmix,son2019qtran,mahajan2019maven}.  In the $m$-step matrix game, zero rewards lead to termination, and the differentiating states are located at the terminal ends. Figure \ref{mstepgame} illustrates the $m$-step matrix game for $m=5$. The optimal policy is to take the top left joint action and finally take the bottom right action, giving an optimal total payoff of  $m+3$.
\begin{figure}[htb!]
	\centering
	\includegraphics[width=0.48\textwidth]{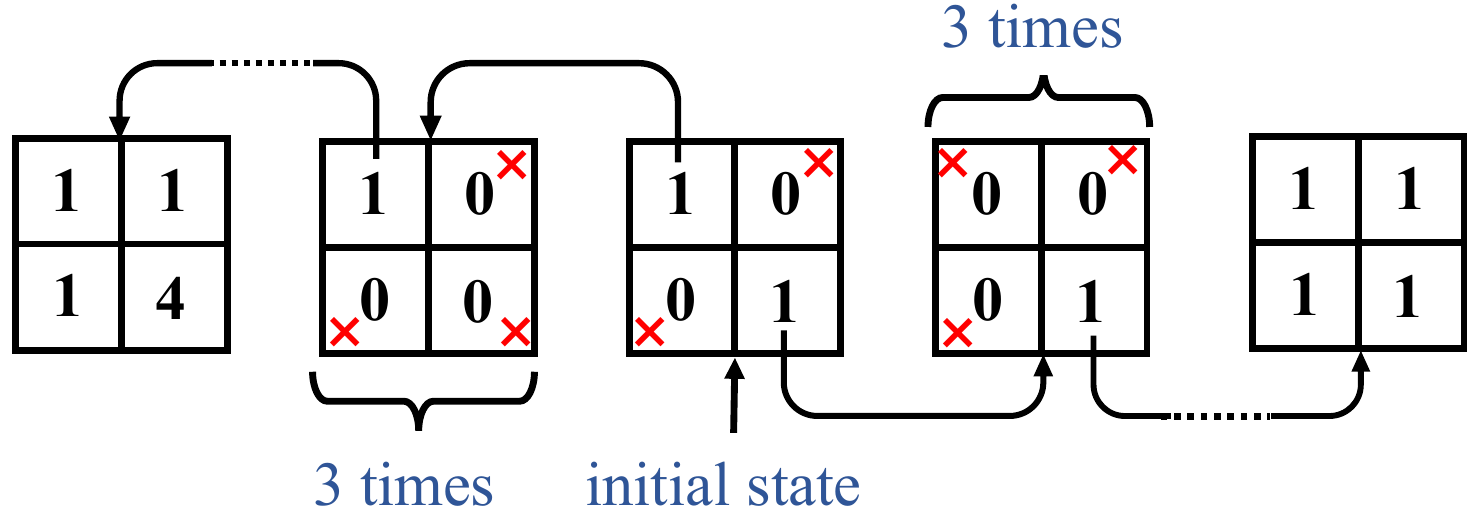}
	\caption{$m$-step matrix game for $m=5$ case.}\label{mstepgame}
\end{figure}
\begin{figure}[htb!]
	\centering
	\includegraphics[width=0.45\textwidth]{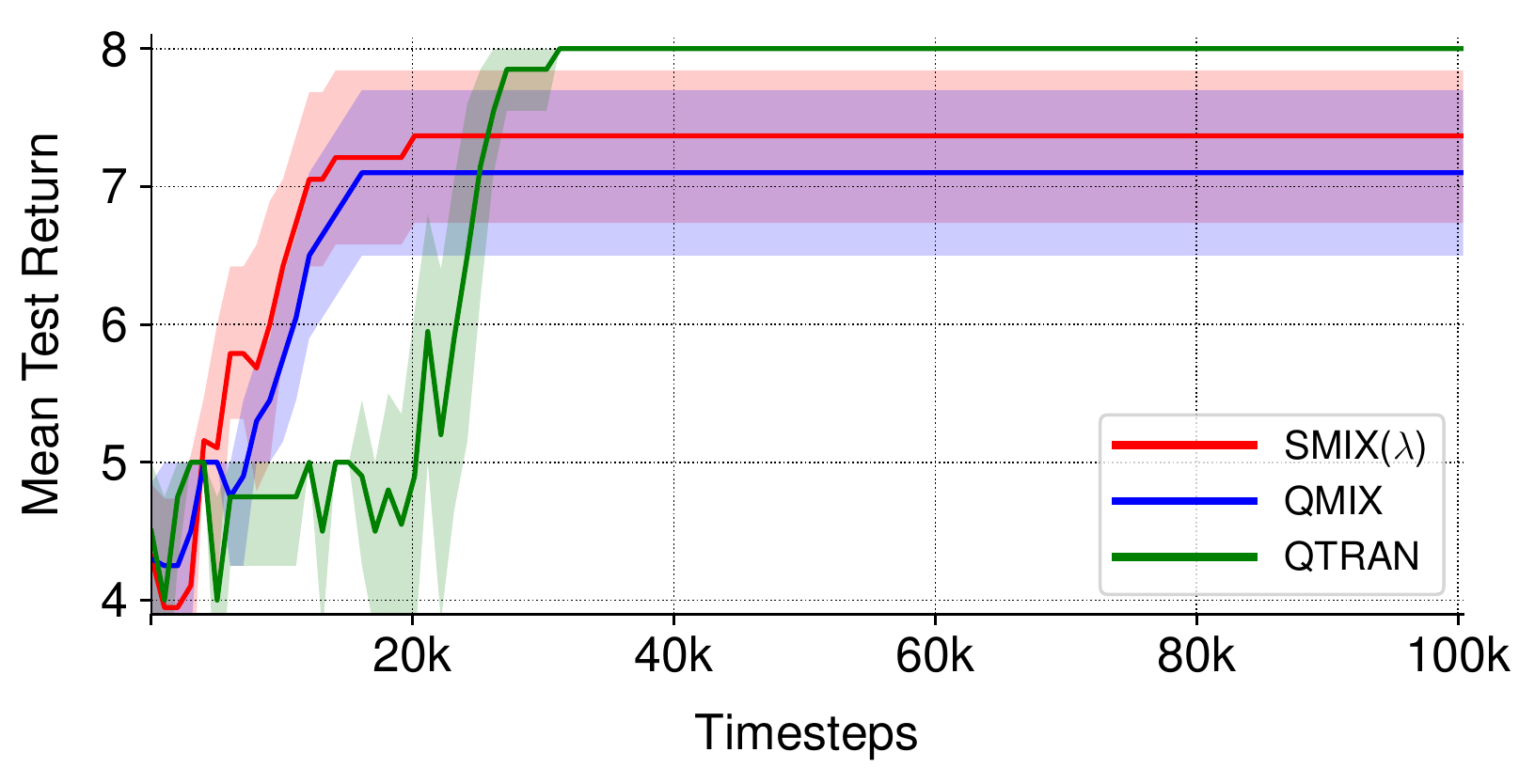}
	\caption{Average return of QMIX, SMIX($\lambda$) and QTRAN on 5-step matrix game for 100k training steps. (The mean and 95\% confidence interval are shown across 20 independent runs.)}\label{n-matrix}
\end{figure}
	
Figure \ref{n-matrix} gives the results on the $m$-step matrix game. One can see that although our SMIX($\lambda$) learns the fast, QTRAN achieves the highest return. This is as expected as QTRAN has the most highest representational complexity among the compared ones, allowing it to achieve the lowest bias in this relatively simple scenario. However, with the increasing complexity of search space (as in SMAC \cite{samvelyan2019starcraft}), a larger hypothesis space could turn out to be a disadvantage as the variance could dominate the generalization error, making the task of finding the best hypothesis quite challenging. In such cases, methods like ours would be a better choice, as illustrated in the experimental Section.}



\section{Theretical Analysis}
\label{sec:analysis}
In this section, we give the convergence analysis of the proposed SMIX($\lambda$) algorithm, by first building the connection between SMIX($\lambda$) and a previous method named $Q(\lambda)$ \cite{harutyunyan2016q}, originally proposed for off-policy value function evaluation under single-agent settings.

Denoting $G^{\boldsymbol{\pi}}$ as the $\lambda$-return estimator (cf., \ref{eq:offpolicyreturn}) for the action-value of the target policy $\boldsymbol{\pi}$, the goal of an off-policy method is to use the data from the behavior policy $\boldsymbol{\mu}$ to correct $G^{\boldsymbol{\pi}}$, in a way such that the following criterion is met,
\begin{equation}
\mathbb{E}_{\boldsymbol{\pi}}\Big[G^{\boldsymbol{\pi}}\Big] = \mathbb{E}_{\boldsymbol{\mu}}\Big[G^{\boldsymbol{\mu},\boldsymbol{\pi}}\Big],\label{eq:offpolicytarget}
\end{equation}
where $G^{\boldsymbol{\mu},\boldsymbol{\pi}}$ is the corrected return of off-policy data.

The most commonly used method for calculating the $G^{\boldsymbol{\mu},\boldsymbol{\pi}}$  is the importance sampling (IS) method which multiply each reward with a weighted term to satisfy \eqref{eq:offpolicytarget}. Indeed, the motivation behind SMIX($\lambda$) is to simplify the IS method so that it can be used in multi-agent settings. If we define the IS ratio at timestep $t$ as: $\rho_t=\frac{\boldsymbol{\pi}(\boldsymbol{a}_t|\boldsymbol{\tau}_t)}{\boldsymbol{\mu}(\boldsymbol{a}_t|\boldsymbol{\tau}_t)}$, then the $n$-step return using IS can be defined as:
\begin{equation}
\begin{aligned}
{{G}}_{t}^{(n)}=& r_{t+1}+\gamma  \rho_{t+1} r_{t+2}+\cdots \\ &+\gamma^{n-1}  \rho_{t+1} \cdots \rho_{t+n-1} r_{t+n}\\ &+\gamma^{n} \rho_{t+1} \cdots \rho_{t+n}\mathbb{E}_{\boldsymbol{\pi}} Q^{\text{SMIX}}(\boldsymbol{\tau}_{t+n},\cdot).
\end{aligned}\label{eq:ISn}
\end{equation}
Thus, we have the following form of $G^{\boldsymbol{\mu},\boldsymbol{\pi}}$:
\begin{equation}
\begin{aligned}
&G^{\boldsymbol{\mu},\boldsymbol{\pi}}= (1-\lambda)\sum_{n=1}^{\infty}\lambda^{n-1}{{G}}_{t}^{(n)}.
\end{aligned}\label{eq:ISlambda}
\end{equation}
Plugging \eqref{eq:ISn} into \eqref{eq:ISlambda}, we have:
\begin{equation}
\begin{aligned}
&G^{\boldsymbol{\mu},\boldsymbol{\pi}}\leftarrow Q^{\text{SMIX}}\left(\boldsymbol{\tau}_{t},\boldsymbol{a}_t\right)+\sum_{k=t}^{\infty}\left(\prod_{i=t+1}^{k} \gamma \lambda \rho_{i}\right){\delta}_k^{\boldsymbol{\pi}},\\
&{\delta}_k^{\boldsymbol{\pi}}=\left(r_{k+1}+\gamma \rho_{k+1} Q^{\text{SMIX}}\left(\boldsymbol{\tau}_{k+1}, \boldsymbol{a}_{k+1}\right)-Q^{\text{SMIX}}\left(\boldsymbol{\tau}_{k}, \boldsymbol{a}_{k}\right)\right).
\end{aligned}\label{eq:smix1}
\end{equation}
In SMIX($\lambda$), all the importance sampling factor are relaxed to $1.0$, then corresponding to (\ref{eq:offpolicyoperator}), the update rule of SMIX($\lambda$) can be expressed as\footnote{We consider the expected form of $Q^{\text{SMIX}}\left(\boldsymbol{\tau}_{k+1}, \boldsymbol{a}_{k+1}\right)$ in \eqref{eq:smix1} and the training data is sampled from a replay buffer.},

\begin{equation}
\begin{small}
\begin{aligned}
&Q^{\textrm{SMIX}}\left(\boldsymbol{\tau}_{t},\boldsymbol{a}_t\right)\leftarrow Q^{\text{SMIX}}\left(\boldsymbol{\tau}_{t},\boldsymbol{a}_t\right)+\mathbb{E}_{\boldsymbol{\mu}}\left[\sum_{k=t}^{\infty}\left(\prod_{i=t+1}^{k}  \lambda\gamma \right){\delta}_k^{\boldsymbol{\pi}}\right],\\
&{\delta}_k^{\boldsymbol{\pi}}=\left(r_{k+1}+\gamma \mathbb{E}_{\boldsymbol{\mu}} Q^{\text{SMIX}}\left(\boldsymbol{\tau}_{k+1}, \cdot\right)-Q^{\text{SMIX}}\left(\boldsymbol{\tau}_{k}, \boldsymbol{a}_{k}\right)\right).
\label{eq:ISupdate}
\end{aligned}
\end{small}
\end{equation}

In contrast with the multiplicative operation for off-policy learning, an additive-type operation is used in $Q(\lambda)$ \cite{harutyunyan2016q}. In particular, an additive correction term $\Delta^{\boldsymbol{\mu},\boldsymbol{\pi}}_r$, is added to each reward when calculating $G^{\boldsymbol{\mu},\boldsymbol{\pi}}$ in \eqref{eq:offpolicytarget} and get:
\begin{equation}
\begin{aligned}
{{G}}_{t}^{(n)}= (r_{t+1}+ \Delta^{\boldsymbol{\mu},\boldsymbol{\pi}}_{r_{t+1}})&+\cdots  +\gamma^{n-1}(r_{t+n}+\Delta^{\boldsymbol{\mu},\boldsymbol{\pi}}_{r_{t+n}}) \\&+\gamma^{n}  \mathbb{E}_{\boldsymbol{\pi}}Q^{Q(\lambda)}(\boldsymbol{\tau}_{t+n},\cdot).
\end{aligned}\label{eq:addterm}
\end{equation}
The major advantage of this additive off-policy correction is that there is no product
of the ratio and no the joint policy $\boldsymbol{\pi}(\boldsymbol{a}|\boldsymbol{\tau})$ involved, hence completely bypassing the limitations of the IS method\footnote{But under the condition that the behavior policy $\boldsymbol{\mu}$ should be close to the target policy $\boldsymbol{\pi}$, which under our experience replay setting should not be a problem (cf., \cite{fujimoto2019off}). }.

Specifically, the updating rule of $Q(\lambda)$ method is \cite{harutyunyan2016q}:
\begin{equation}
\begin{small}
\begin{aligned}
&Q^{Q(\lambda)}(\boldsymbol{\tau}_t, \boldsymbol{a}_t)\leftarrow Q^{Q(\lambda)}(\boldsymbol{\tau}_t, \boldsymbol{a}_t)+\mathbb{E}_{\boldsymbol{\mu}}\left[\sum_{k=t}^{\infty}\left(\prod_{i=t+1}^{k}\lambda \gamma\right) \hat{\delta}_{k}^{\boldsymbol{\pi}}\right],\\
&\hat{\delta}_k^{\boldsymbol{\pi}}=\left(r_{k+1}+\Delta^{\boldsymbol{\mu},\boldsymbol{\pi}} r_{k+1}\right),\\&\Delta^{\boldsymbol{\mu},\boldsymbol{\pi}} r_{k+1}=\gamma \mathbb{E}_{\boldsymbol{\pi}} Q^{Q(\lambda)}\left(\boldsymbol{\tau}_{k+1}, \cdot\right)-Q^{Q(\lambda)}\left(\boldsymbol{\tau}_{k}, \boldsymbol{a}_{k}\right).\label{eq:qlambda}
\end{aligned}
\end{small}
\end{equation}
By comparing (\ref{eq:ISupdate}) and (\ref{eq:qlambda}), we see that our SMIX($\lambda$) and off-policy $Q(\lambda)$ are essentially equivalent except that SMIX($\lambda$) calculates $\mathbb{E}_{\boldsymbol{\mu}} Q^{\text{SMIX}}\left(\boldsymbol{\tau}_{k+1}, \cdot\right)$ in ${\delta}_{k}^{\boldsymbol{\pi}}$ while $Q(\lambda)$ calculate $\mathbb{E}_{\boldsymbol{\pi}} Q^{Q(\lambda)}\left(\boldsymbol{\tau}_{k+1}, \cdot\right)$ in $\hat{\delta}_{k}^{\boldsymbol{\pi}}$.

The following theorem states that when $\boldsymbol{\pi}$ and $\boldsymbol{\mu}$ are sufficiently close, the difference between the output of SMIX($\lambda$) and $Q(\lambda)$ is bounded. This implies that SMIX($\lambda$) is consistent with the $Q(\lambda)$ algorithm.
\begin{Theo}\label{theo:matching}
	Suppose we update the value function from $Q^{\text{SMIX}}_n(\boldsymbol{\tau}_{t},\boldsymbol{a}_{t})=Q^{Q(\lambda)}_n(\boldsymbol{\tau}_{t},\boldsymbol{a}_{t})$, where $n$ represents the $n$-th update. Let $\epsilon=\max_{\boldsymbol{\tau}}\|\boldsymbol{\pi}(\cdot|\boldsymbol{\tau})-\boldsymbol{\mu}(\cdot|\boldsymbol{\tau})\|_1$, $M=\max_{\boldsymbol{\tau},\boldsymbol{a}}|Q^{Q(\lambda)}_{n}(\boldsymbol{\tau},\boldsymbol{a})|$. Then, the error between $Q^{\text{SMIX}}_{n+1}(\boldsymbol{\tau}_{t},\boldsymbol{a}_t)$ and $Q^{Q(\lambda)}_{n+1}(\boldsymbol{\tau}_{t},\boldsymbol{a}_t)$ can be bounded by the expression:
	\begin{equation}
	| Q^{\text{SMIX}}_{n+1}\left(\boldsymbol{\tau}_{t},\boldsymbol{a}_t\right) -  Q^{Q(\lambda)}_{n+1}\left(\boldsymbol{\tau}_{t}, \boldsymbol{a}_t\right)|\leq \frac{\epsilon\gamma}{1-\lambda\gamma} M.\label{eq:theorem2}
	\end{equation}
\end{Theo}
\begin{IEEEproof}
	First, we have,
	\begin{equation*}
	\begin{aligned}
	\left|\delta_{t}^{\boldsymbol{\pi}} - \hat{\delta}_{t}^{\boldsymbol{\pi}}\right| &=\left| \gamma\mathbb{E}_{\boldsymbol{\mu}} Q^{\text{SMIX}}_{n}\left(\boldsymbol{\tau}_{t+1}, \cdot\right)-\gamma\mathbb{E}_{\boldsymbol{\pi}} {Q}^{Q(\lambda)}_{n}\left(\boldsymbol{\tau}_{t+1}, \cdot\right)\right|\\
	&=\gamma\left|\sum_{\boldsymbol{a}}\boldsymbol{\mu}({\boldsymbol{a}}|\boldsymbol{\tau}_{t+1})Q^{\text{SMIX}}_{n}\left(\boldsymbol{\tau}_{t+1}, \cdot\right)-\right.\\&\left.\quad\quad\quad\quad\quad\quad\sum_{\boldsymbol{a}}\boldsymbol{\pi}({\boldsymbol{a}}|\boldsymbol{\tau}_{t+1}){Q}^{Q(\lambda)}_{n}\left(\boldsymbol{\tau}_{t+1}, \cdot\right)\right|\\&\leq \gamma\epsilon M.
	\end{aligned}
	\end{equation*}
	Thus,
	\begin{equation*}
	\begin{aligned}
	&\left| Q^{\text{SMIX}}_{n+1}\left(\boldsymbol{\tau}_{t},\boldsymbol{a}_t\right) -  Q^{Q(\lambda)}_{n+1}\left(\boldsymbol{\tau}_{t}, \boldsymbol{a}_t\right)\right|\\&= \left|\mathbb{E}_{\boldsymbol{\mu}}\left[\sum_{k=t}^{\infty}\left(\prod_{i=t+1}^{k}  \lambda\gamma \right){\delta}_k^{\boldsymbol{\pi}}\right] - \mathbb{E}_{\boldsymbol{\mu}}\left[\sum_{k=t}^{\infty}\left(\prod_{i=t+1}^{k}\lambda \gamma\right) \hat{\delta}_{k}^{\boldsymbol{\pi}}\right]\right|\\
	&=\left|\mathbb{E}_{\boldsymbol{\mu}}\left[\sum_{k=t}^{\infty}\left(\prod_{i=t+1}^{k}  \lambda\gamma \right)\left({\delta}_k^{\boldsymbol{\pi}}-\hat{\delta}_k^{\boldsymbol{\pi}}\right)\right]\right|\\&\leq \left|\mathbb{E}_{\boldsymbol{\mu}}\left[\sum_{k=t}^{\infty}\left(\prod_{i=t+1}^{k}  \lambda\gamma \right)\left(\gamma\epsilon M\right)\right]\right|\\
	& \leq \mathbb{E}_{\boldsymbol{\mu}}\left[\frac{1}{1-\lambda\gamma}(\gamma\epsilon M)\right] = \frac{\epsilon\gamma }{1-\lambda\gamma}M.
	\end{aligned}
	\end{equation*}
	Therefore, the expression (\ref{eq:theorem2}) holds.\qedhere
\end{IEEEproof}
{This theorem indicates that SMIX($\lambda$) has the similar convergence property to $Q(\lambda)$ when the difference between behavior policy $\boldsymbol{\mu}$ and target policy $\boldsymbol{\pi}$ is bounded by $\epsilon$, which is $\epsilon=\max_{\boldsymbol{\tau}}\|\boldsymbol{\pi}(\cdot|\boldsymbol{\tau})-\boldsymbol{\mu}(\cdot|\boldsymbol{\tau})\|_1$. In practice, this
condition can be easily implemented by periodically replacing the current behavior policy with the old
target policy and limiting the size of the replay memory. The following theorem presents the convergence property of the $Q(\lambda)$ method \cite{harutyunyan2016q}.}

\begin{Theo}\cite{harutyunyan2016q}
	Consider the sequence of Q-functions computed according to (\ref{eq:qlambda}) with fixed policy $\boldsymbol{\pi}$ and $\boldsymbol{\mu}$. Let $\epsilon=\max_{\boldsymbol{\tau}}\|\boldsymbol{\pi}(\cdot|\boldsymbol{\tau})-\boldsymbol{\mu}(\cdot|\boldsymbol{\tau})\|_1$. If $\lambda\epsilon<\frac{1-\gamma}{\gamma}$, then under the following conditions:
	\begin{itemize}
		\item
		$\sum_{t\geq 0}\mathbb{P}\{\boldsymbol{\tau}_t,\boldsymbol{a}_t=\boldsymbol{\tau},\boldsymbol{a}\}\geq D > 0.$, where $\mathbb{P}(\boldsymbol{\tau},\boldsymbol{a})$ represents the visit frequency,
		\item
		$\mathbb{E}_{\boldsymbol{\mu}_n}T_n^2<\infty$, where $T_n$ is the length of $\boldsymbol{\tau}_n,$
		\item
		$\sum_{n\geq 0}\alpha_n(\boldsymbol{\tau},\boldsymbol{a}) = \infty, \sum_{n\geq 0}\alpha^2_n(\boldsymbol{\tau},\boldsymbol{a})<\infty$, where $\alpha_n$ is the step-size of the $n$-th iteration,
	\end{itemize}
	we have, almost surely:
	\begin{equation*}
	\lim_{n\rightarrow\infty}Q^{Q(\lambda)}_{n}(\boldsymbol{\tau},\boldsymbol{a}) = Q^{\boldsymbol{\pi}}(\boldsymbol{\tau},\boldsymbol{a}).
	\end{equation*}\label{theo:consistency}
\end{Theo}

The above analysis shows that SMIX($\lambda$) and $Q(\lambda)$ have similarities both formally and analytically. However, when applying them to the problem of multi-agent reinforcement learning, their computational complexity is fundamentally different. This is because to calculate the additive error correction term, $Q(\lambda)$ has to  estimate the expectation over target policy $\boldsymbol{\pi}$ in (\ref{eq:qlambda}), but this is unrealistic in the multi-agent setting  since the dimension of the joint action space grows exponentially with the number of agents. By contrast, the SMIX($\lambda$) relies on the experience replay technique to compute the expectation in (\ref{eq:ISupdate}), whose computational complexity grows only  linearly with the number of training samples, regardless of the size of joint action space and the number of agents involved. Such  scalability makes our method more appropriate for the task of multi-agent reinforcement learning, compared to the  $Q(\lambda)$ and QTRAN (which suffers from the same problem as  $Q(\lambda)$).

Finally, before ending this section, we summarize some of the key characteristics of QMIX and SMIX($\lambda$) in Table \ref{tab:comparision}.

\begin{table}[h!]
	\centering
	\caption{The comparison of QMIX and SMIX($\lambda$).}
	\resizebox{\columnwidth}{!}{
		{\begin{tabular}{|c||c||c|}
			\hline
			\textbf{Property} & \textbf{QMIX} & \textbf{SMIX}($\lambda$) \\
			\hline\hline
			Uses experience replay & \cmark & \cmark \\
			\hline
			CVF estimation & Q-learning based & Expected-SARSA based \\
			\hline
			TD target & one-step return & $\lambda$-return \\
			\hline
			Algorithm's Flexibility & CGB assumption & No explicit CGB\\
			\hline
			Stable point of convergence & $Q^{*}$ & $Q^{*}$ \\
			\hline
	\end{tabular}}}
	\label{tab:comparision}
\end{table}
\begin{table*}[htb!]
	\caption{The scenarios considered in our experiments.}
	\begin{center}
		\resizebox{\textwidth}{!}{
			\begin{tabular}{|c|| c||c||c|}
				\hline
				{\makecell*[c]{Name}}& Ally Units & Enemy Units&Type\\                                    \hline
				\hline
				3m & 3 Marines &3 Marines & homogeneous \& symmetric\\
				\hline
				8m & 8 Marines &8 Marines& homogeneous \& symmetric\\
				\hline
				2s3z & 2 Stalkers \& 3 Zealots & 2 Stalkers \& 3 Zealots&heterogeneous \& symmetric\\
				\hline
				3s5z & 3 Stalkers \& 5 Zealots & 3 Stalkers \& 5 Zealots&heterogeneous \& symmetric\\
				\hline
				2m\_vs\_1z & 2 Marines & 1 Zealot&asymmetric\\
				\hline
				2s\_vs\_1sc& 2 Stalkers & 1 Spine Crawler&asymmetric\\
				\hline
				3s\_vs\_3z& 3 Stalkers & 3 Zealots&asymmetric\\
				\hline
				1c3s5z&1 Colossi, 3 Stalkers \& 5 Zealots &1 Colossi, 3 Stalkers \& 5 Zealots&heterogeneous \& symmetric\\
				\hline
				MMM& 1 Medivac, 2 Marauders \& 7 Marines& 1 Medivac, 2 Marauders \& 7 Marines&heterogeneous \& symmetric\\
				\hline
			\end{tabular}
		}		
	\end{center}
	\label{table:scenarios}
\end{table*}
\section{Experiments}
\label{sec:experiments}

In this section, we first describe the environmental setup and the implementation details of our method. Then we give the experimental results and ablation study. {The code of SMIX($\lambda$) is available at: https://github.com/chaovven/SMIX.}

\subsection{Environmental Setup}

We evaluate our SMIX($\lambda$) in the StarCraft Multi-Agent Challenge (SMAC) \cite{samvelyan2019starcraft} environment. The  SMAC is chosen as our testbed mainly because of the following two reasons: (1) SMAC provides a set of rich cooperative scenarios that challenge algorithms to handle significant partial observability and credit assignment problem \cite{foerster2018counterfactual}. These problems bring a great challenge for centralized value function estimation. (2) SMAC also provides an open-source Python-based implementation of several key algorithms, which allows for fair comparisons between different methods.

SMAC is based on the popular real-time strategy (RTS) game StarCraft II\footnote{StarCraft II is a trademark of Blizzard Entertainment$^\text{TM}$.}. Each unit can be seen as an individual agent which has a complex set of micro-actions. Different from the full StarCraft II game, SMAC focuses on fully cooperative, decentralized \emph{micromanagement} multi-agent problems. \emph{Micromanagement} means the task of controlling individual or grouped units to fight enemy units. While high-level strategies such as economy and resource management, known as \emph{macromanagement}, are not considered in SMAC.

SMAC provides several challenging micro scenarios that aim to evaluate different aspects of cooperative behaviors of a group of agents. In each scenario, two groups of agents are placed on the map with random initial positions within groups at the beginning of each episode. Each agent can only receive local observations within its \emph{sight range}, which introduces significant partial observability. Extra global state information is available during centralized training. The units of the first group (allied units) are controlled by decentralized agents, while the units of the other group (enemy units) are controlled by built-in heuristic game AI bot with difficulty ranging from \emph{very easy} to \emph{cheat insane}. In our experiments, we set the difficulty of the game AI bot to \emph{very difficult} for all experiments. Available actions for each agent contain: \texttt{move[direction]}, \texttt{attack[enemy id]}, \texttt{stop}, and \texttt{noop}. Agents receive a joint reward equal to the total damage dealt on the enemy units. We use the default setting for the reward. Refer to \cite{samvelyan2019starcraft} for more details.

\begin{figure}[htb!]
	\centering
	\subfloat[3 Staklers vs 5 Zealots]{\includegraphics[width=0.24\textwidth]{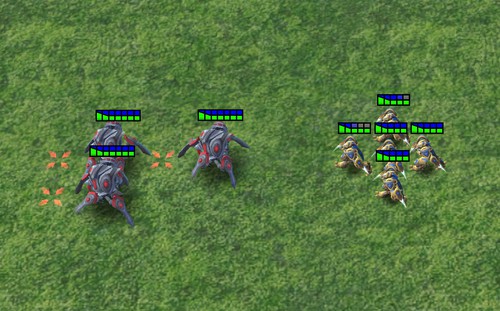}}\hfill
	\subfloat[8 Marines vs 8 Marines]{\includegraphics[width=0.24\textwidth]{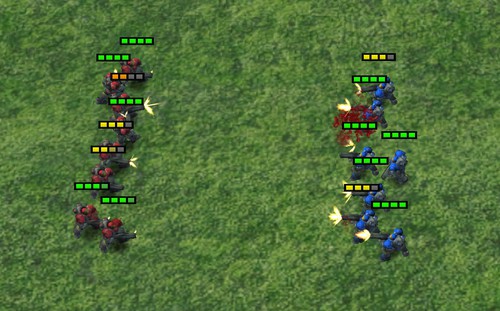}}
	\caption{Screenshots of two SMAC scenarios.}
	\label{fig:scenarios}
\end{figure}
\begin{figure*}[h!]
	\centering
	\subfloat[3m]{\includegraphics[width=0.33\textwidth]{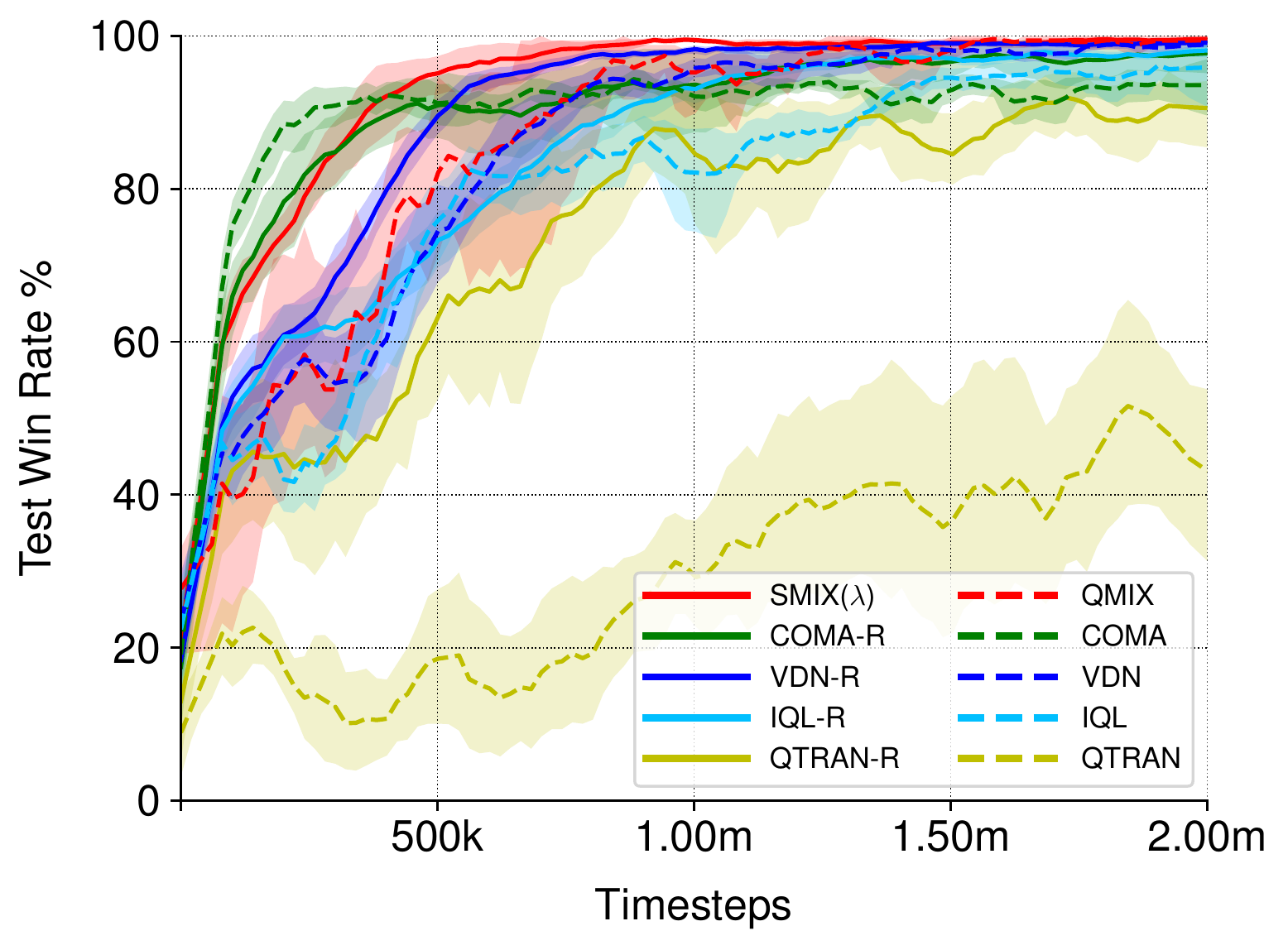}}
	\subfloat[8m]{\includegraphics[width=0.33\textwidth]{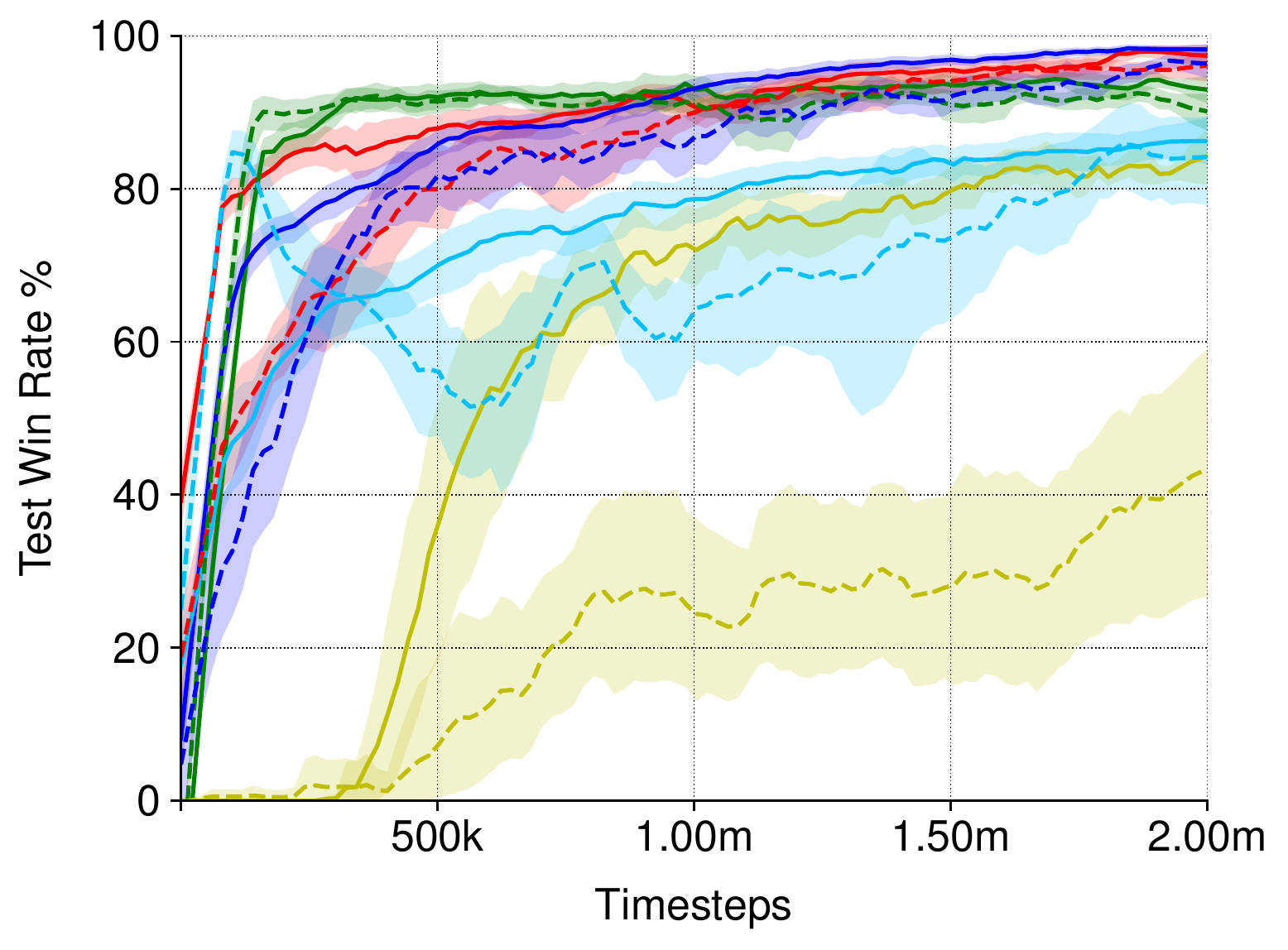}}
	\subfloat[2s3z]{\includegraphics[width=0.33\textwidth]{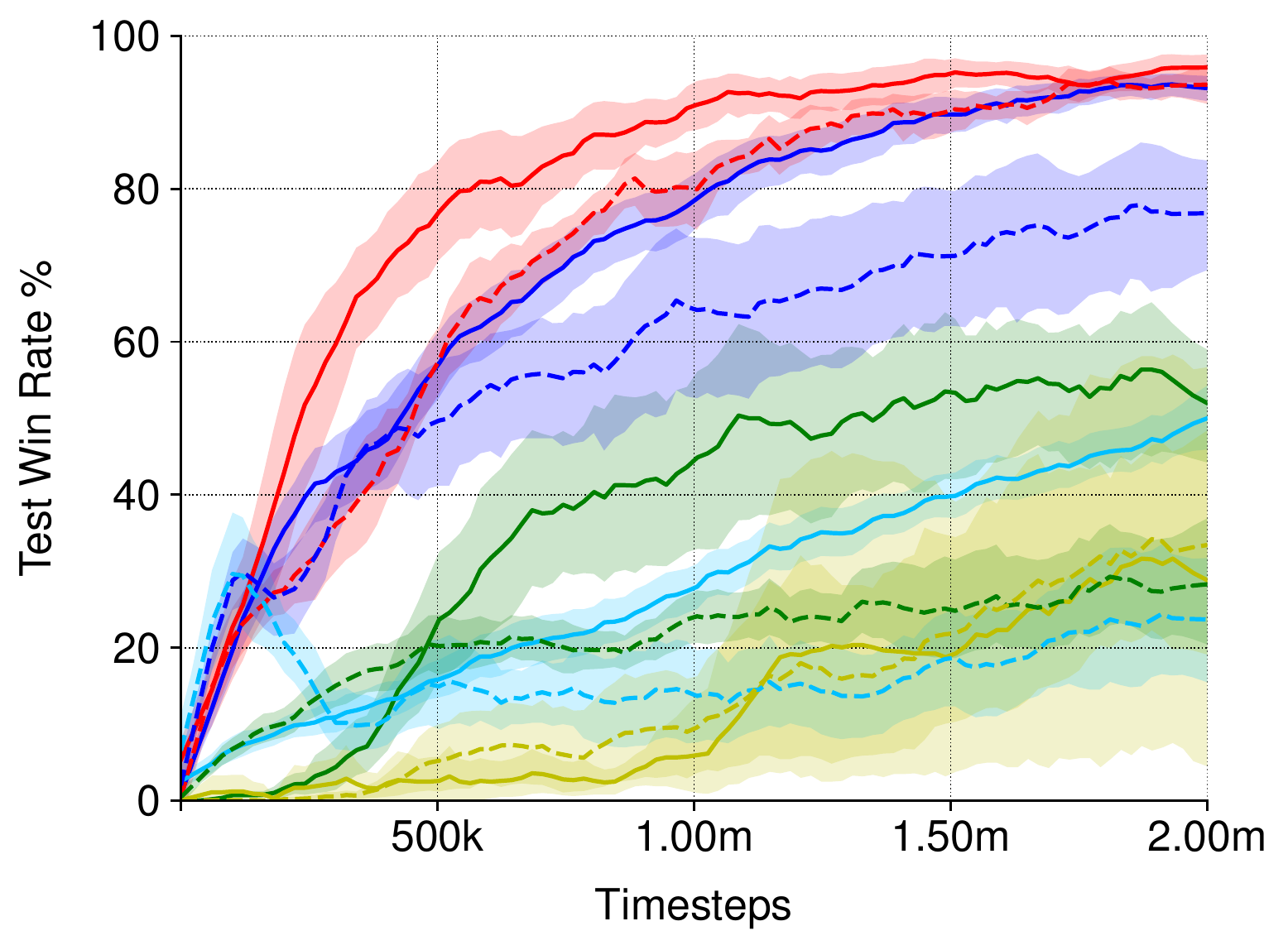}}\\
	\subfloat[3s5z]{\includegraphics[width=0.33\textwidth]{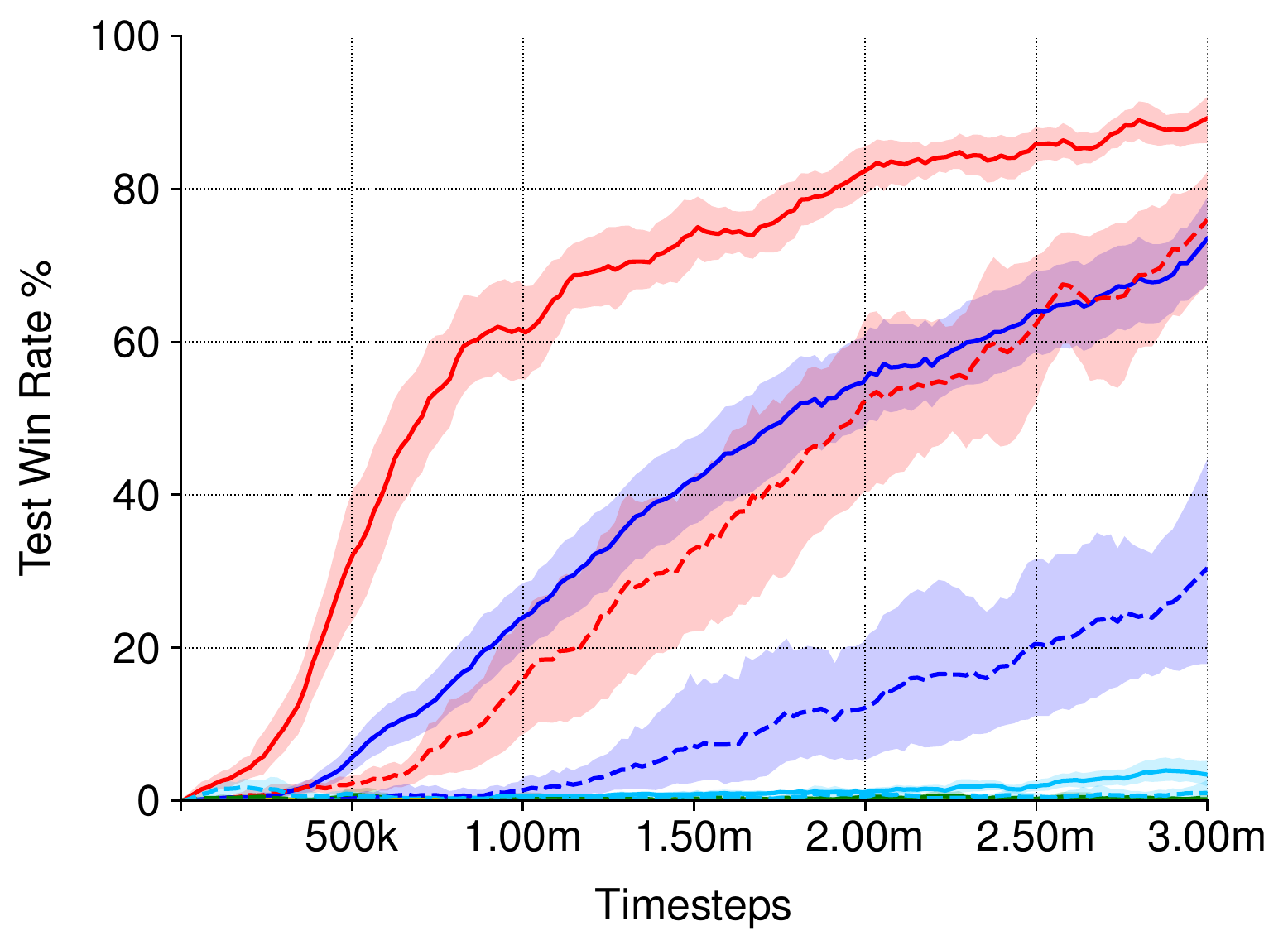}}
	\subfloat[2m\_vs\_1z]{\includegraphics[width=0.33\textwidth]{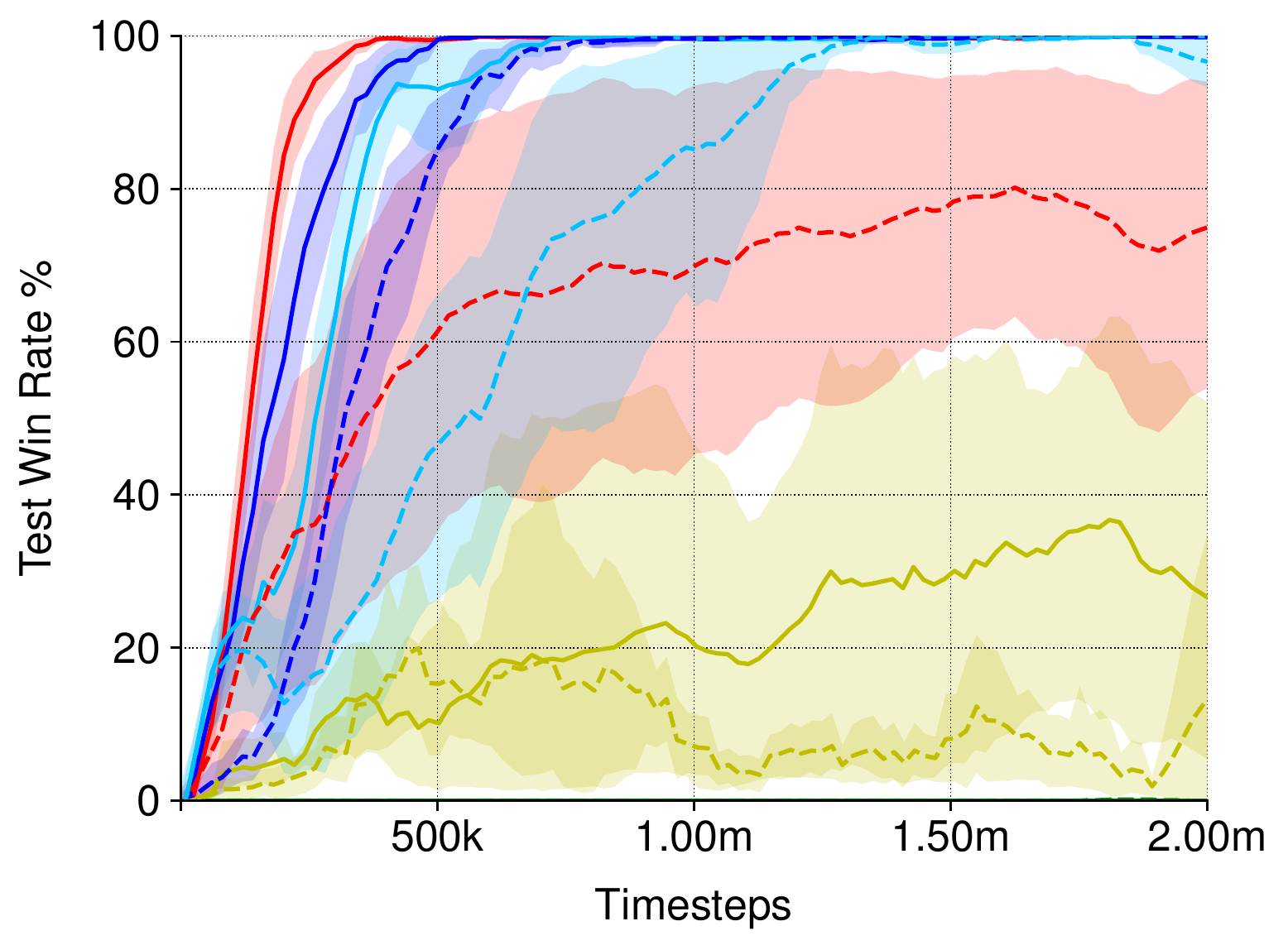}}
	\subfloat[2s\_vs\_1sc]{\includegraphics[width=0.33\textwidth]{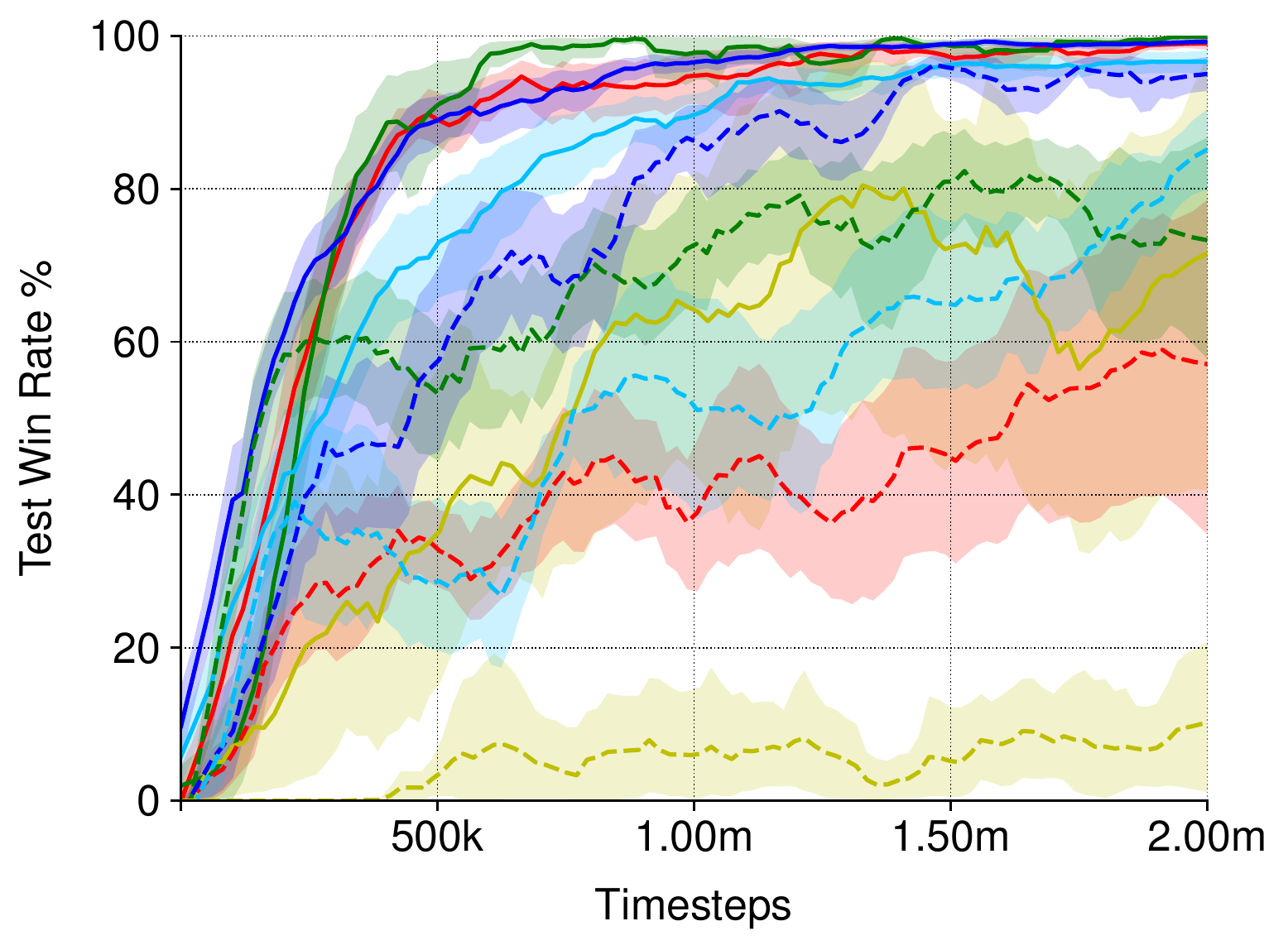}}\\
	\subfloat[3s\_vs\_3z]{\includegraphics[width=0.33\textwidth]{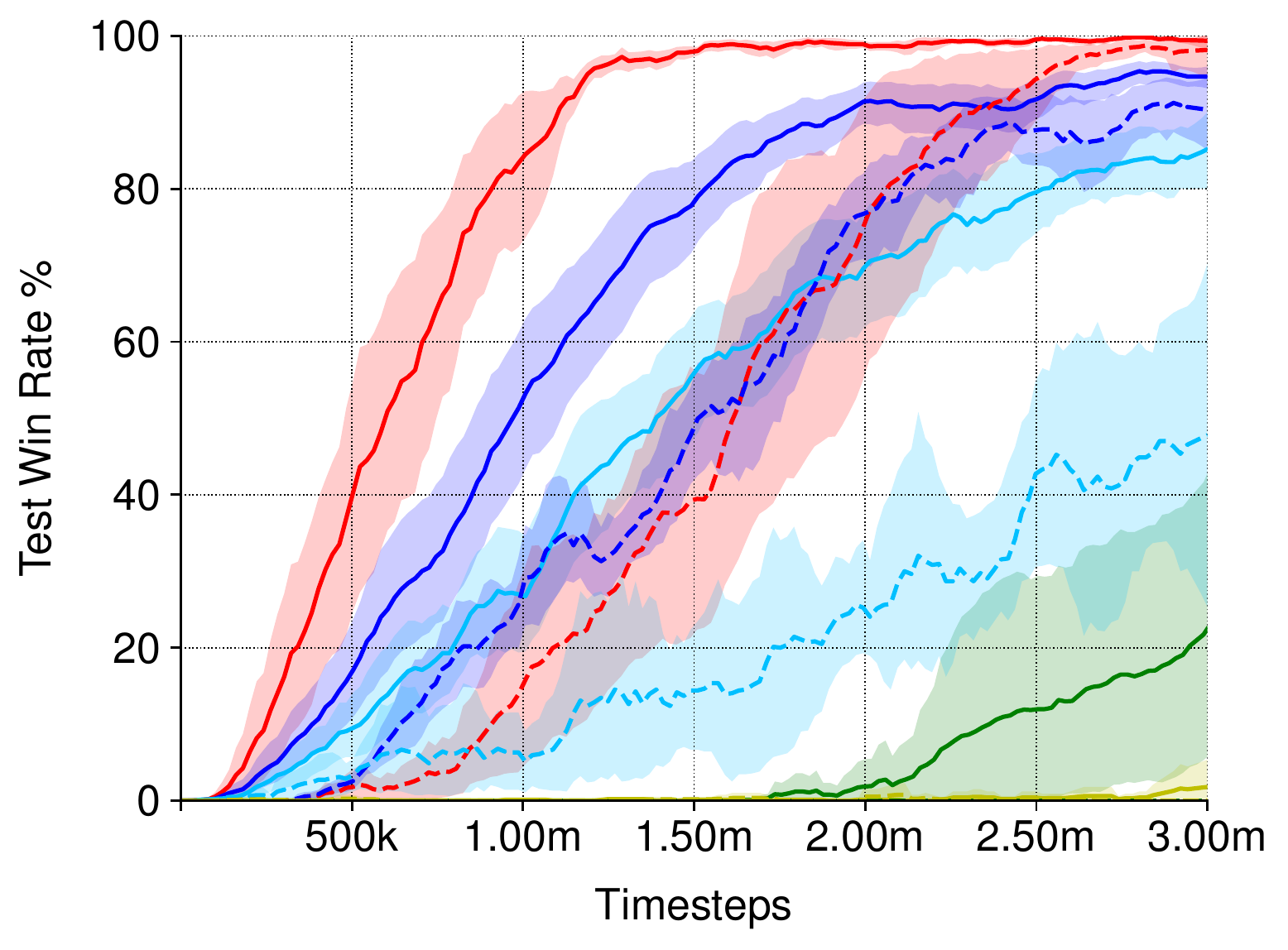}}
	\subfloat[1c3s5z]{\includegraphics[width=0.33\textwidth]{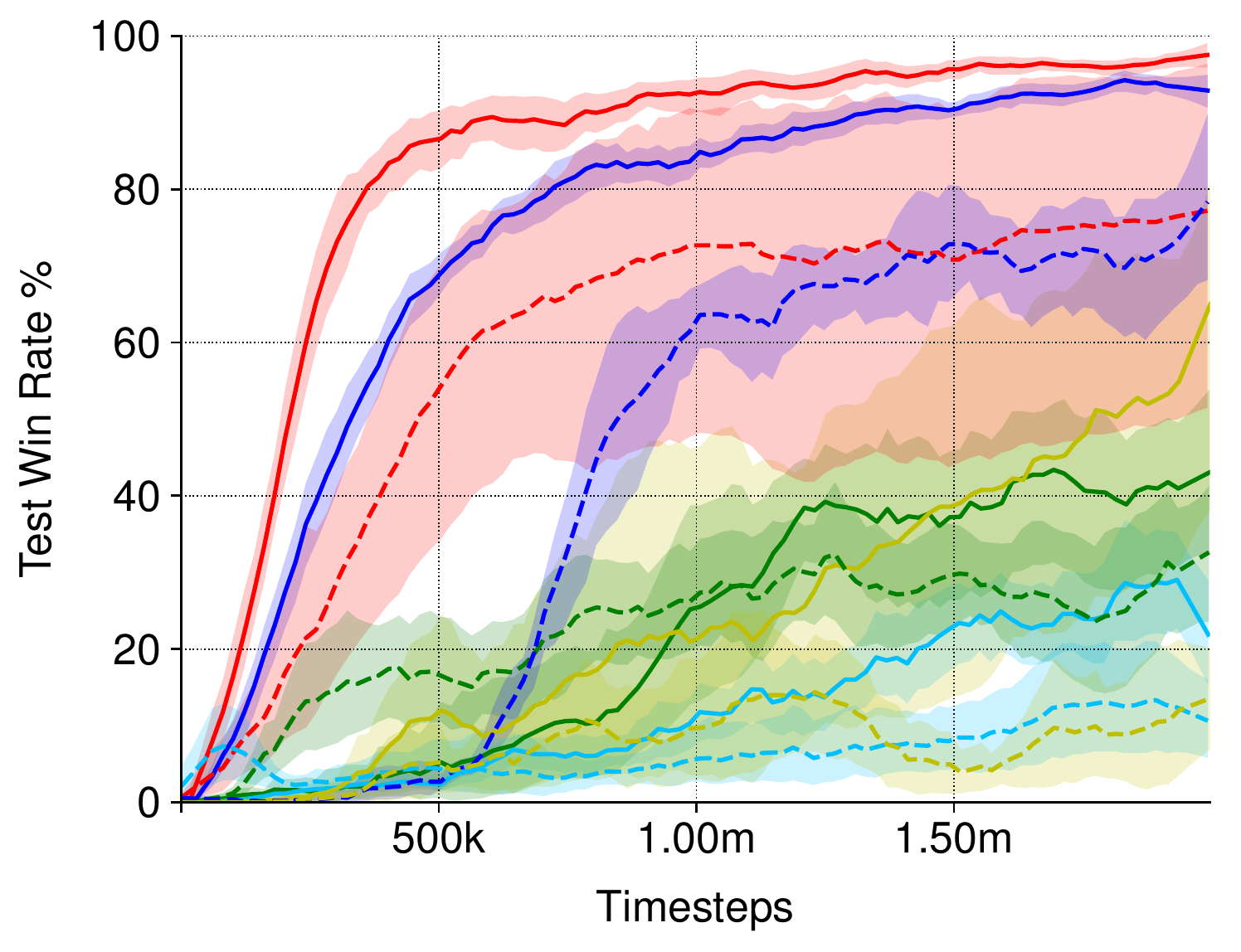}}
	\subfloat[MMM]{\includegraphics[width=0.33\textwidth]{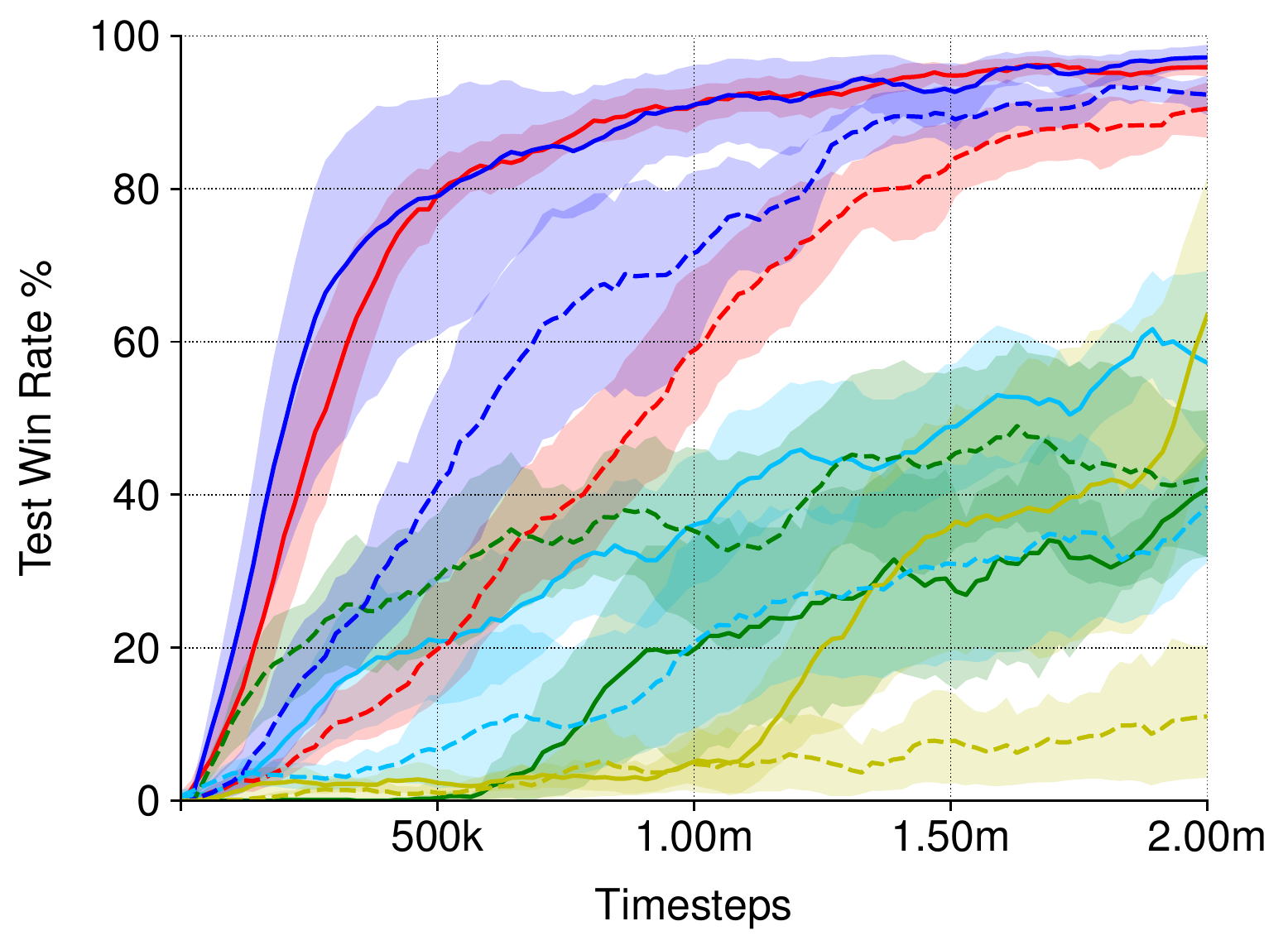}}
	\caption{{Test win rates for SMIX($\lambda$), revised methods (COMA-R, VDN-R, IQL-R, QTRAN-R) and comparison methods (QMIX, COMA, VDN, IQL, QTRAN) in nine different scenarios. Revised methods are created by replacing original methods' CVF estimation procedures with the one adopted by SMIX($\lambda$). The performance of our method and revised methods and plotted with solid line, with their counterparts shown with dashed lines of the same color. The mean and 95\% confidence interval are shown across 10 independent runs. The legend in (a) applies across all plots.
	}}
	\label{fig:main_results}
\end{figure*}

The following 3 types of scenarios are considered in our experiments:
(1) \emph{homogeneous and symmetric units},
(2) \emph{heterogeneous and symmetric units},
(3) \emph{asymmetric units}. The list of scenarios considered in our experiment is presented in Table \ref{table:scenarios}. Figure \ref{fig:scenarios} shows the screenshots of two SMAC scenarios used in our experiments.

We use \emph{test win rate} as the evaluation metric, which is proposed in \cite{samvelyan2019starcraft} and is the default evaluation metric in the SMAC environment. The test win rate is evaluated in the following procedure: the training process is interrupted after every 20,000 timesteps, then 24 independent test episodes are run with each agent performing greedy action selection in a decentralized way. Test win rate refers to the percentage of episodes where the agents defeat all enemy units within the time limit.

\subsection{Implementation Details}
\label{sec:imple_detail}
The agent network architecture of SMIX($\lambda$) consists of a 64-dimensional GRU \cite{GRU}. One 64-dimensional fully connected layer with ReLU activation function before GRU is applied for processing the input. The layer after GRU is a fully connected layer of 64 units, which outputs the decentralized action-values $Q^i(\tau, \cdot)$ of agent $i$. All agent networks share parameters for reducing the number of parameters to be learned. Thus the agent's one-hot index $i$ is concatenated onto each agent's observations. The agent's previous action is also concatenated to the input.

Based on the basic network architecture of QMIX \cite{rashid2018qmix}, SMIX($\lambda$) performs the centralized value function estimation with $\lambda$-return ($\lambda$ = 0.8) calculated from a batch of 32 episodes. The batch is sampled uniformly from a replay buffer that stores the most recent 1500 episodes. We run 4 episodes simultaneously. 
Then we perform training on those fully unrolled episodes. The target network is updated after every 200 training episodes. $\lambda$ is set to 0.8. 

The $\epsilon$-greedy method is used in the training procedure for exploration. $\epsilon$ is annealed linearly from 1.0 to 0.05 across the first 50k timesteps for all experiments. The discount factor $\gamma$ is set to 0.99, and the RMSprop optimizer is used with learning rate  $lr = 0.0005$ and $\alpha=0.99$ without weight decay or momentum during training.



\subsection{Comparative Evaluation}
We compare our SMIX($\lambda$) with state-of-the-art algorithms QMIX \cite{rashid2018qmix} and COMA \cite{foerster2018counterfactual}, which currently perform the best on the SMAC benchmark. VDN \cite{sunehag2018value} and IQL \cite{tan1993multi} are chosen as baselines for comparisons. {QTRAN \cite{son2019qtran} is also implemented in SMAC for comprehensive comparison.}

The results of all methods in the training process are plotted in Figure \ref{fig:main_results} and we also provide quantitative comparisons of our methods and their counterparts after training for 1 million steps in Table \ref{table:quantitative_res}. Overall, SMIX($\lambda$) significantly outperforms all the comparison methods in heterogeneous or asymmetric scenarios (i.e., scenarios except 3m and 8m), while performing comparably to them in homogeneous and symmetric scenarios (i.e., 3m and 8m) both in terms of the learning speed and final performance. 

In homogeneous and symmetric scenarios such as 3m and 8m, COMA is only slightly faster than SMIX($\lambda$) but underperforms SMIX($\lambda$) in terms of the final performance. In asymmetric (e.g., 3s\_vs\_3z, 2s\_vs\_1sc) or heterogeneous (e.g., 2s3z, 3s5z, 1c3s5z) maps, COMA fails to solve these scenarios effectively. {It's worth noting that QTRAN does not work very well in these complex environments, which is also consistent with the experimental results of other authors \cite{mahajan2019maven}. One possible reason is that QTRAN needs to address a very large optimization problem in the high-dimensional joint action space. This highlights the importance of improving the performance in value function estimation.}

Due to the poor performance of COMA and QTRAN, QMIX can be seen as the state-of-the-art on this benchmark. However, the learning speed of SMIX($\lambda$) is almost twice as fast as QMIX. In 3s5z, SMIX($\lambda$) (solid red line) achieves a nearly 90\% win rate, while the best comparison method QMIX (dotted red line) achieves about only 70\% test win rate.
In 2s\_vs\_1sc, SMIX($\lambda$) also requires less than half the number of samples of QMIX and other comparison methods to reach the asymptotic performance. The largest performance gap can be seen in 3s\_vs\_3z map (Figure \ref{fig:main_results}g). QMIX needs to be trained for nearly 3 million timesteps to achieve a 100\% test win rate, while half of the timesteps are sufficient for SMIX($\lambda$) to achieve the same win rate. In MMM, we can find an interesting result that the VDN can achieve better performance than QMIX (see Figure \ref{fig:main_results}i). This indicates that a simpler network structure can also have enough representative capacity and the reason for VDN's superior performance is that a simpler network architecture only needs a relatively small number of samples for training. Furthermore, by incorporating the proposed centralized training method, VDN's performance can be further improved, which implies that the bottleneck of VDN and QMIX may be the bias and variance of the CVF estimation. On the whole, the superior performance of SMIX($\lambda$) using $\lambda$-return with off-policy episodes presents a clear benefit over the one-step estimation of QMIX.

\renewcommand\cellgape{\Gape[0.5pt]}
\setcellgapes{8pt}
\begin{table*}[h!]
	\caption{Mean, standard deviation, and median of test win rate percentages after training for 1 million timesteps in nine different scenarios.}
	\begin{center}
		\resizebox{\textwidth}{!}{
			\begin{tabular}{|c| c||c |c||c| c||c| c||c|c||c|c|}
				\hline
				\multicolumn{2}{|c||}{\makecell*[c]{Algorithms}}               &
				SMIX($\lambda$)                & QMIX         &
				COMA-R              & COMA       &
				VDN-R        & VDN &IQL-R & IQL & {QTRAN-R }& {QTRAN}                                                                                                                                                                                                             \\
				\hline
				\hline
				\multirow{2}{*}{3m} & \multicolumn{1}{c||}
				{mean $\pm$ \emph{std}}& {\textbf{99}} ($\pm$0) & 95 ($\pm$3) & \textbf{93 ($\pm$8)}   & 92 ($\pm$2)  & \textbf{98} ($\pm$0)&  95 ($\pm$2) & \textbf{91} ($\pm$4)  & 83 ($\pm$9)& {\textbf{84} ($\pm$10)}& {{29} ($\pm$11)}\\
				\cline{2-12}
				&  median & \makecell[c]{\textbf{99}} & 95 & \textbf{97} & 93 & \textbf{98} & 95 &\textbf{94} & 86 & {\textbf{86}}& {{29}} \\
				
				\hline
				\multirow{2}*{8m} & \multicolumn{1}{c||}
				{mean $\pm$ \emph{std}}& \makecell[c]{\textbf{91} ($\pm$3)} & 90 ($\pm$3) & \textbf{92} ($\pm$2)   & 90 ($\pm$2)  & \textbf{94} ($\pm$3)&  86 ($\pm$5) & \textbf{80} ($\pm$5)  & 59 ($\pm$15)& {\textbf{72} ($\pm$6)}& {{24} ($\pm$17)}\\
				\cline{2-12}
				&  median & \makecell[c]{\textbf{90}} & 89 & \textbf{93} & 91 & \textbf{93} & 87 &\textbf{79} & 58 & {\textbf{72}}& {{21}}\\
				
				\hline
				\multirow{2}*{2s3z} & \multicolumn{1}{c||}
				{mean $\pm$ \emph{std}}& \makecell[c]{\textbf{90}}($\pm$4) & 81 ($\pm$7) & \textbf{44} ($\pm$18)   & 24 ($\pm$6)  & \textbf{78} ($\pm$14)&  64 ($\pm$16) & \textbf{32} ($\pm$8)  & 14 ($\pm$10)& {{5} ($\pm$8)}& {\textbf{9} ($\pm$10)}\\
				\cline{2-12}
				&  median & \makecell[c]{\textbf{91}} & 81 & \textbf{47} & 24 & \textbf{79} & 71 &\textbf{31} & 13 & {{2}}& {\textbf{5}}\\
				
				\hline
				\multirow{2}*{3s5z} & \multicolumn{1}{c||}
				{mean $\pm$ \emph{std}}& \makecell[c]{\textbf{61} ($\pm$11)} & 16 ($\pm$12) & \textbf{0} ($\pm$0)   & \textbf{0} ($\pm$0)  & \textbf{29} ($\pm$12)&  1 ($\pm$2) & \textbf{0} ($\pm$0)  & \textbf{0} ($\pm$0)& {\textbf{0} ($\pm$0)}& {\textbf{0} ($\pm$0)}\\
				\cline{2-12}
				&  median & \makecell[c]{\textbf{62}} & 11 & \textbf{0} & \textbf{0} & \textbf{26} & 0 &\textbf{0} & \textbf{0}&{\textbf{0}}&{\textbf{0}} \\
				\hline
				\multirow{2}*{2m\_vs\_1z} & \multicolumn{1}{c||}
				{mean $\pm$ \emph{std}}&\makecell[c]{\textbf{99}}($\pm$0) & 69 ($\pm$38) & \textbf{0} ($\pm$0)   & \textbf{0} ($\pm$0)  & \textbf{99} ($\pm$0)&  \textbf{99} ($\pm$0) & \textbf{99} ($\pm$0)  &85 ($\pm$27)& {\textbf{20} ($\pm$29)} & {{7} ($\pm$4)}\\
				\cline{2-12}
				&  median & $\textbf{100}$ & 99 & \textbf{0} & \textbf{0} & \textbf{99} & \textbf{99} &\textbf{100} & {99} & {\textbf{4}}& {{6}}\\
				\hline
				\multirow{2}*{2s\_vs\_1sc} & \multicolumn{1}{c||}
				{mean $\pm$ \emph{std}}& \makecell[c]{\textbf{94} ($\pm$5)} & 39 ($\pm$19) & \textbf{97} ($\pm$4)   & 77 ($\pm$11)  & \textbf{96} ($\pm$2)&  86 ($\pm$8) & \textbf{92} ($\pm$6)  & 51 ($\pm$22)&{\textbf{63} ($\pm$24)} & {{6} ($\pm$11)}\\
				\cline{2-12}
				&  median & \makecell[c]{\textbf{96}} & 45 & \textbf{100} & 78 & \textbf{97} & 88 &\textbf{94} & 54 &{\textbf{70}}& {{0}}\\
				\hline
				\multirow{2}*{3s\_vs\_3z} & \multicolumn{1}{c||}
				{mean $\pm$ \emph{std}}&\makecell[c]{\textbf{84}} ($\pm$14) & 15 ($\pm$20) & \textbf{0} ($\pm$0)   & \textbf{0} ($\pm$0)  & \textbf{67} ($\pm$25)&  27 ($\pm$9) & \textbf{35} ($\pm$21)  & 5 ($\pm$4)&{\textbf{0} ($\pm$0)}& {{0} ($\pm$0)}\\
				\cline{2-12}
				&  median & $\textbf{88}$ & 9 & \textbf{0} & \textbf{0} & \textbf{83} & {27} &\textbf{31} & 6 &{\textbf{0}}& {{0}} \\
				\hline
				\multirow{2}*{1c3s5z} & \multicolumn{1}{c||}
				{mean $\pm$ \emph{std}}&\makecell[c]{\textbf{92}} ($\pm$3) & 72 ($\pm$32) & {24} ($\pm$19)   & \textbf{27} ($\pm$13)  & \textbf{84} ($\pm$3)&  61 ($\pm$5) & \textbf{11} ($\pm$5)  & 5 ($\pm$6) & {\textbf{21} ($\pm$29)}& {{9} ($\pm$10)}\\
				\cline{2-12}
				&  median & $\textbf{93}$ & 88 & {18} & \textbf{28} & \textbf{85} & {61} &\textbf{11} & 3 & {\textbf{9}}& {{8}}\\
				\hline
				\multirow{2}*{MMM} & \multicolumn{1}{c||}
				{mean $\pm$ \emph{std}}&\makecell[c]{\textbf{91}} ($\pm$4) & 59 ($\pm$15) & {20} ($\pm$17)   & \textbf{34} ($\pm$18)  & \textbf{91} ($\pm$9)&  72 ($\pm$17) & \textbf{35} ($\pm$14)  & 20 ($\pm$19) &{\textbf{5} ($\pm$4)}& {{4} ($\pm$5)}\\
				\cline{2-12}
				&  median & $\textbf{91}$ & 61 & {22} & \textbf{39} & \textbf{94} & {78} &\textbf{34} & 15 & {{2}} & {\textbf{3}} \\
				\hline
			\end{tabular}
		}		
	\end{center}
	\label{table:quantitative_res}
\end{table*}

\begin{figure*}[htb!]
	\centering
	\subfloat[$\lambda$ for SMIX($\lambda$) (3s5z)\label{fig_lambda_3s5z}]{\includegraphics[width=0.325\textwidth]{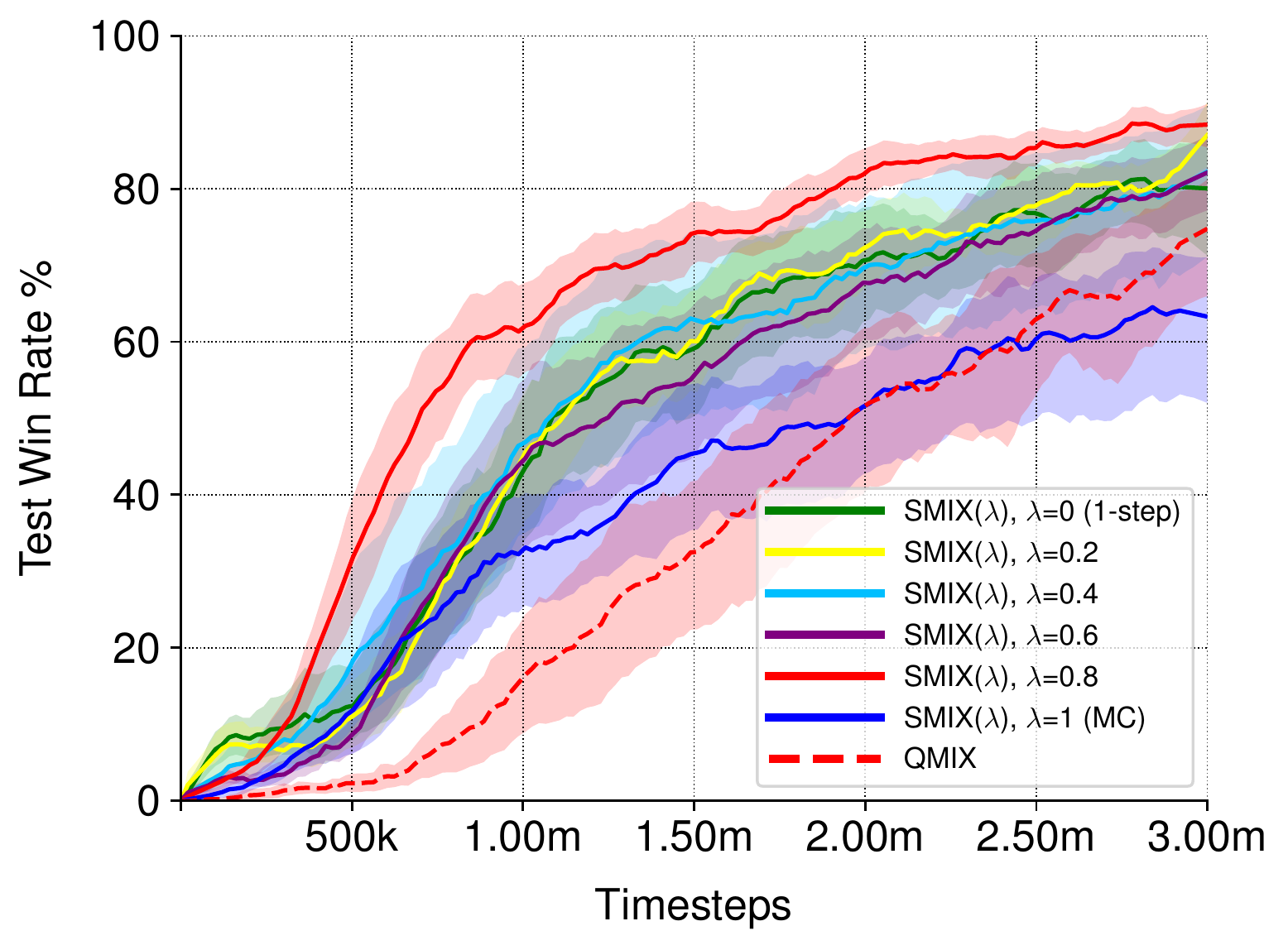}}
	\subfloat[Step $n$ for $n$-step SMIX (3s5z)\label{fig_n_3s5z}]{\includegraphics[width=0.325\textwidth]{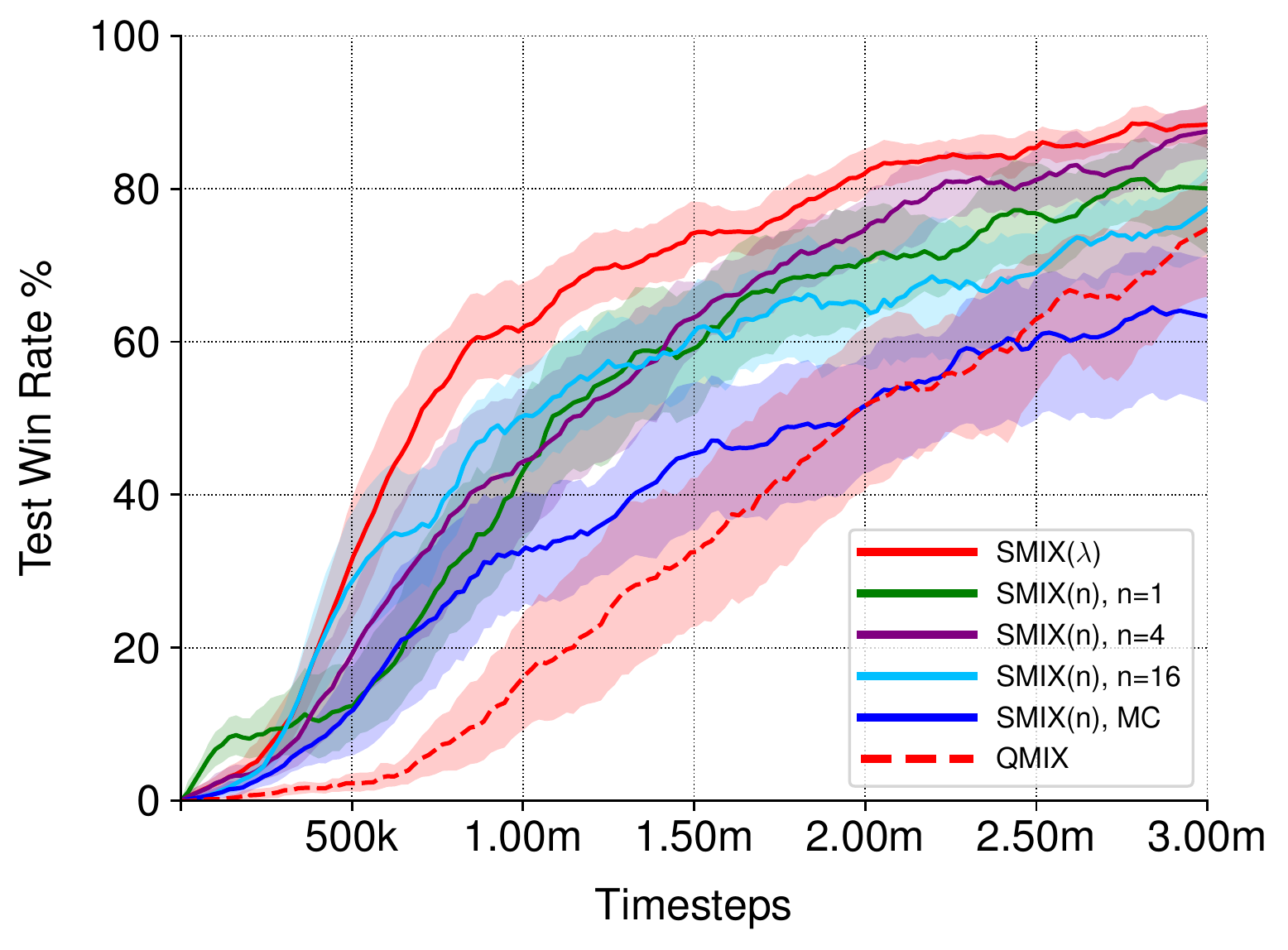}}
	\subfloat[Buffer size $b$ for SMIX($\lambda$) (3s5z)\label{fig_buffer_3s5z}]{\includegraphics[width=0.325\textwidth]{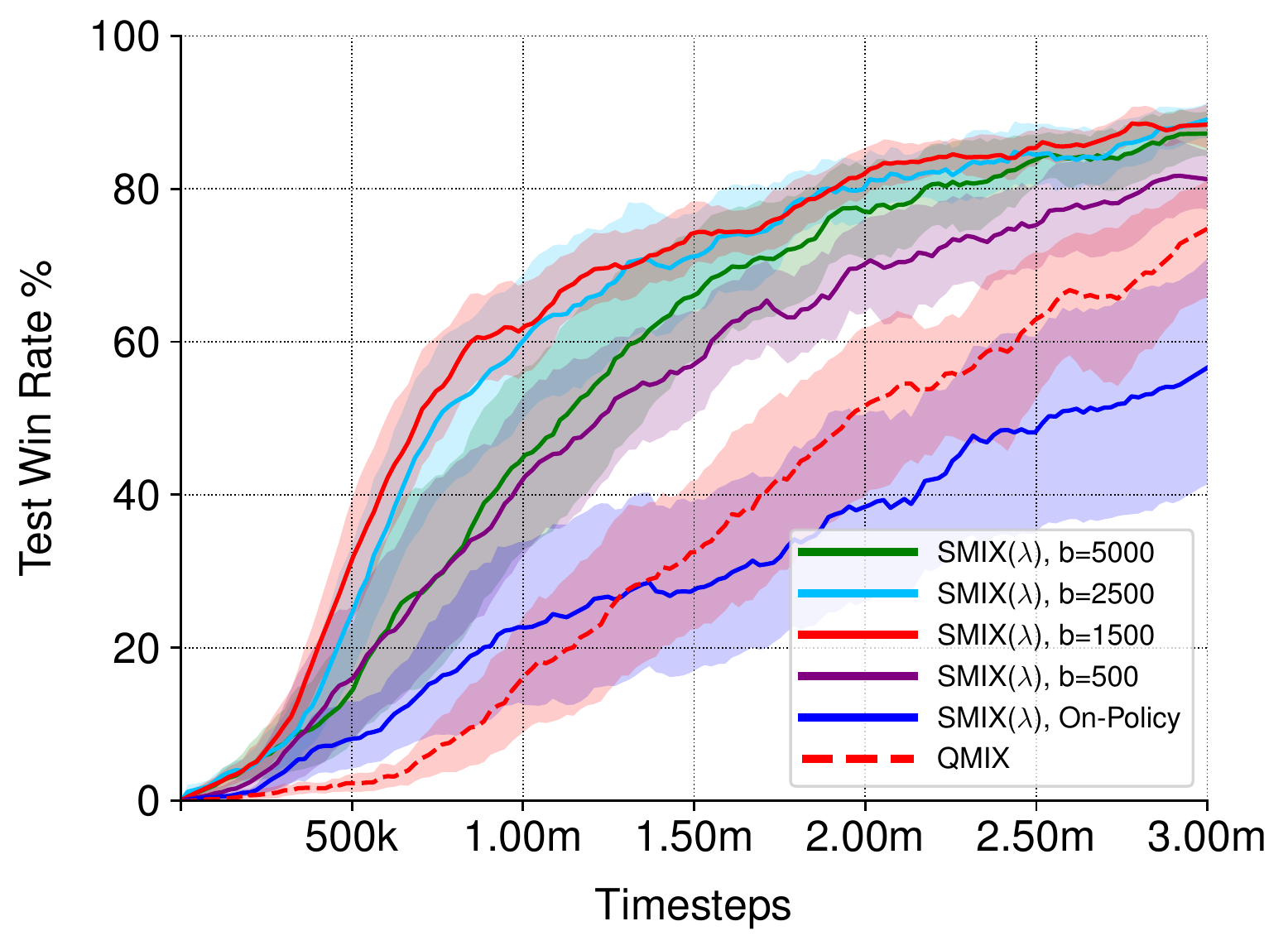}}\\
	\subfloat[$\lambda$ for SMIX($\lambda$) (2s\_vs\_1sc)\label{fig_lambda_2svs1sc}]{\includegraphics[width=0.325\textwidth]{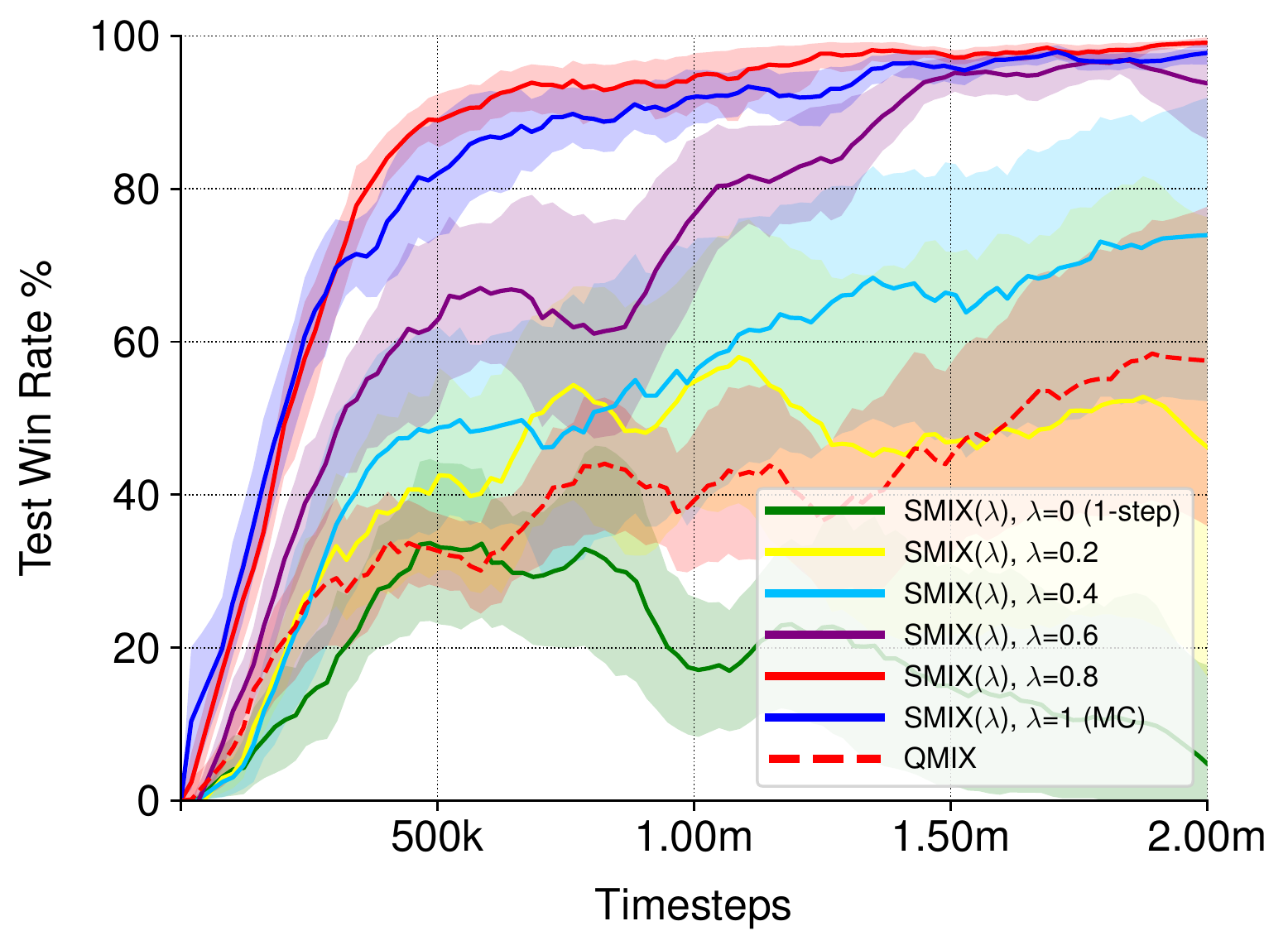}}
	\subfloat[Step $n$ for $n$-step SMIX (2s\_vs\_1sc)\label{fig_n_2svs1sc}]{\includegraphics[width=0.325\textwidth]{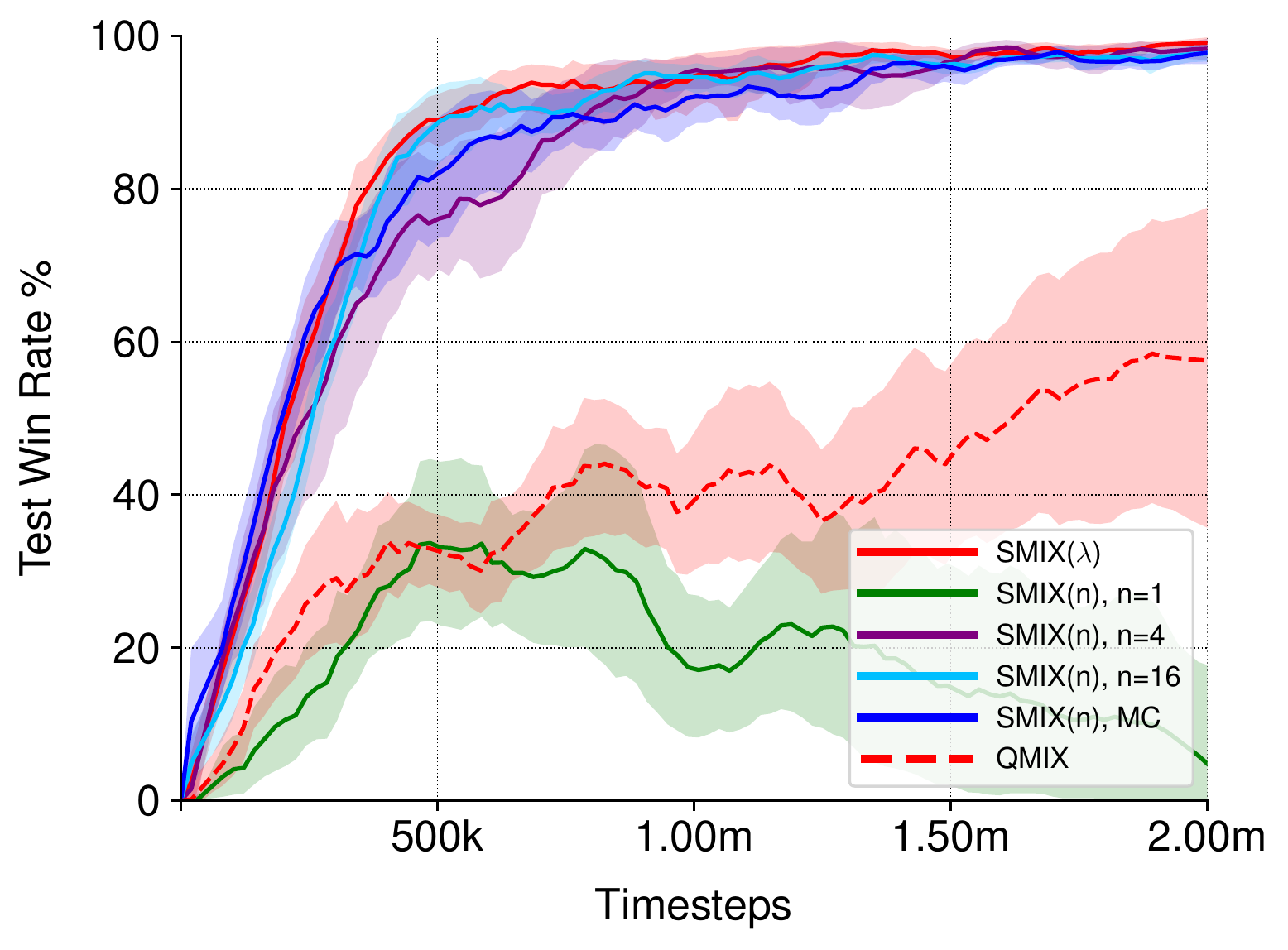}}
	\subfloat[Buffer size $b$ for SMIX($\lambda$) (2s\_vs\_1sc)\label{fig_buffer_2svs1sc}]{\includegraphics[width=0.325\textwidth]{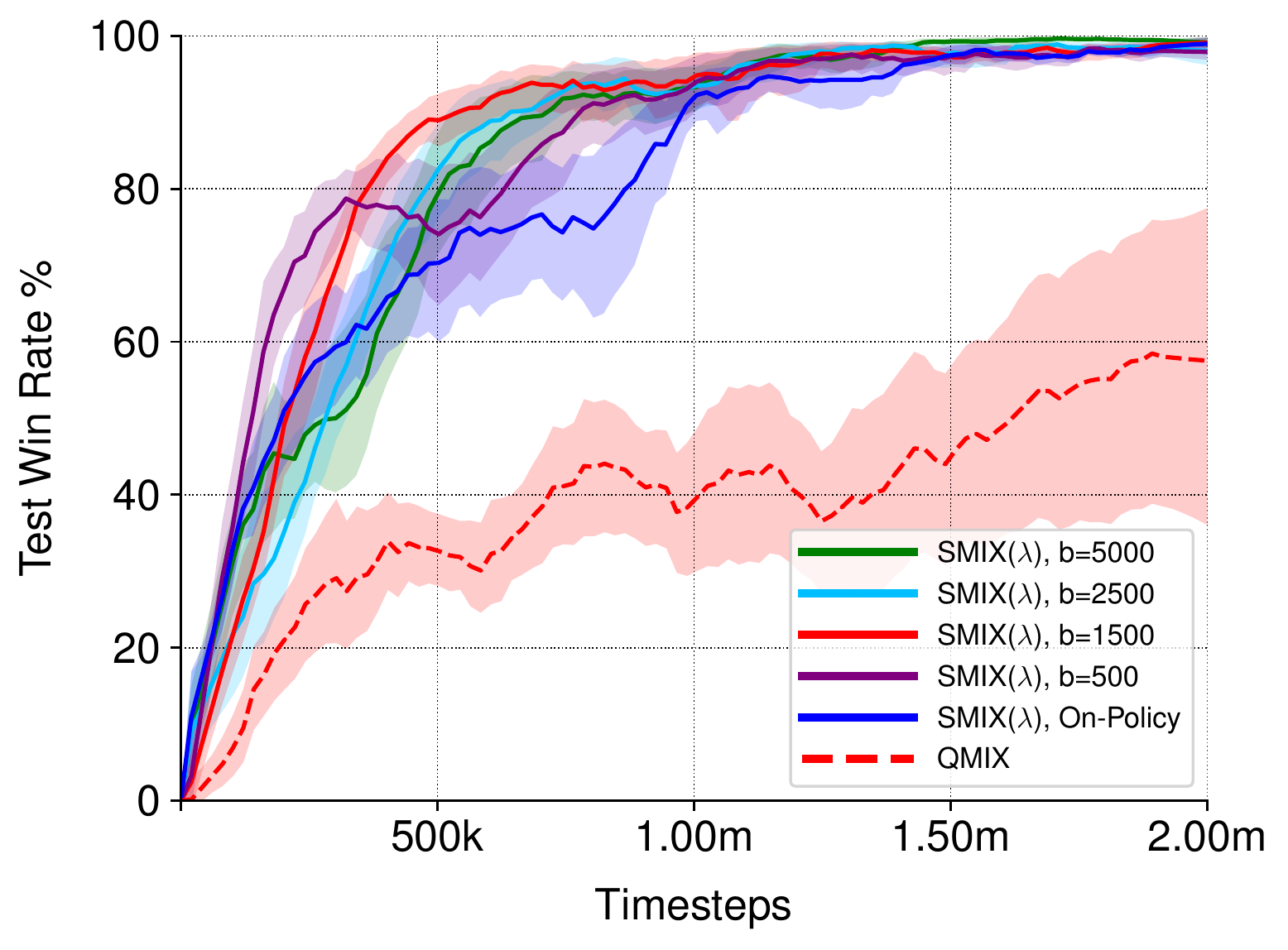}}
	\caption{
		Sensitivity of SMIX($\lambda$) to selected hyperparameters in two different scenarios.
		The mean and 95\% confidence interval is shown across 10 independent runs. The performance of the baseline (QMIX) is shown as a dashed red line.
		(a) and (d) show the sensitivity of SMIX($\lambda$) to the value of $\lambda$;
		(b) and (e) show the results using n-step TD with different backup steps;
		(c) and (f) show the comparison between SMIX($\lambda$) and its on-policy version.
	}
	\label{fig:ablation}
\end{figure*}

\subsection{Generalizing SMIX($\lambda$) to Other MARL Algorithms}

SMIX($\lambda$) focuses on centralized value function evaluation with $\lambda$-return calculated from off-policy episodes. This method could ideally be generalized to other MARL algorithms incorporating critic estimation, including critic-only and actor-critic algorithms.

To demonstrate the benefits of our approach, we generalize the CVF estimation procedure of SMIX($\lambda$) to the following algorithms: COMA, VDN, IQL and QTRAN.
{We achieve these by replacing their original value function estimation procedure with ours (see Section \ref{sec:methods}), yielding four revised algorithms called COMA-R, VDN-R, IQL-R and QTRAN-R respectively.}
Figure \ref{fig:main_results} gives the comparisons between our methods and their counterparts.
Overall, most of the extended methods (solid line) perform on par or significantly better than their counterparts (in the same color but dashed line) in most scenarios both in terms of the final win rate and learning speed.

VDN-R considerably improves the performance of VDN. Especially under challenging scenarios such as 3s5z, VDN-R achieves about 75\% final win rate, which is more than twice as that of VDN (nearly 30\%). Such improvement may be contributed to $\lambda$-return and the independence of the unrealistic centralized greedy assumption during the learning phase.

Furthermore, we find that VDN-R performs even better than QMIX in most scenarios. Recall that VDN uses a linear combination of decentralized Q-values (and so does our VDN-R) and QMIX extends VDN by combining decentralized Q-values in a non-linear way. Thus, QMIX can represent a richer class of CVF than VDN. However, our results indicate that the performance bottleneck of VDN may not be the limited representational capacity, but how to effectively balance the bias and variance in the estimation of CVF.

{Similarly, by comparing QTRAN-R and QTRAN in Figure \ref{fig:main_results}, one can see significant performance improvement by replacing QTRAN's CVF with ours, indicating that the CVF estimation plays an essential role in the QTRAN system and our proposed method is beneficial.}

Also, COMA-R improves COMA's performance in most scenarios, which can be considered as a success of utilizing the off-policy data, as COMA also adopts $\lambda$-return but uses only the on-policy data.

Another observation is that our method also works for IQL, which is a fully decentralized MARL algorithm. This suggests that our method is not limited to \emph{centralized} value function estimation but also applicable to \emph{decentralized} cases.

It is worth mentioning that the extended methods may not make improvements if the original methods do not work, e.g., QTRAN, COMA, IQL, and their counterparts do not work in 3s5z scenario (see Figure \ref{fig:main_results}d and Table \ref{table:quantitative_res}).
The reason may be that the main limitations of QTRAN, COMA and IQL on 3s5z do not lie in the inaccurate value function estimation, but rather in other problems, e.g., scaling not well to a large number of agents and multi-agent credit assignment problem.

{\subsection{Sample Efficiency Analysis}
	\label{subsec:sample_efficiency}
To study the sample efficiency of SMIX($\lambda$), we fix the amount of samples used in training procedure to compare the sample efficiency of different algorithms. The quantitative comparisons of different methods after training for 1 million steps (1 million interactions with the environment) are presented in Table \ref{table:quantitative_res}.

Overall, SMIX($\lambda$) achieves the best performance consistently across all scenarios among compared algorithms, after being trained from the same number of interaction with the environment. This can be partially explained by the fact that our algorithm is off-policy by design. In other words, it has the capability to learn from the previous experience/policies, hence improving the sample efficiency.}

\subsection{Ablation Study}
\label{sec:ablation}

We perform the ablation experiments to investigate the necessity of balancing the bias and variance and the influence of utilizing the off-policy data.

\noindent
\textbf{$\boldsymbol{\lambda}$-Return vs. n-Step Returns.}
To investigate the necessity of balancing the bias and variance in multi-agent problems, we adjust the parameter $\lambda$, where larger $\lambda$ corresponds to smaller bias and larger variance whereas smaller $\lambda$ indicates the opposite. Especially, $\lambda=0$ is equivalent to \emph{one-step return} (corresponding to the largest bias and the smallest variance); $\lambda=1$ is equivalent to \emph{Monte-Carlo (MC) return} ($\infty$-step, corresponding to the smallest bias and the largest variance). We also evaluate a variant named SMIX($n$), which uses n-step return in place of $\lambda$-return as the TD target, i.e., $y_{t}^{tot} = \sum_{i=1}^{n} \gamma^{i-1} r_{t+i} + \gamma^n Q(\boldsymbol{\tau}_{t+n}, \boldsymbol{a}_{t+n}; \theta^-)$.

As Figure \ref{fig_lambda_3s5z} and \ref{fig_lambda_2svs1sc} show, SMIX($\lambda$) with $\lambda=0.8$ consistently achieves the best performance in selected scenarios.
The method with $\lambda=1$ (MC return, blue line) performs the worst in 3s5z, while $\lambda=0$ (one-step return, green line) performs the worst in 2s\_vs\_1sc.
These results reveal that the large variance of MC return or large bias of one-step return may degrade the performance.
Similar results could also be seen in SMIX($n$) (Figure \ref{fig_n_3s5z} and \ref{fig_n_2svs1sc}), where SMIX($n$) with $n=4$ performs the best in 3s5z, while the one with $n=16$ performs the best in 2s\_vs\_1sc. It is not easy to find the same $n$ for SMIX($n$) as SMIX($\lambda$) which sets $\lambda=0.8$ and performs consistently well across different maps. {These results are consistent with the results in literature \cite{fedus2020revisiting} that multi-step returns are beneficial for experience replay based algorithms.}
In summary, it is necessary to balance the trade-off between bias and variance in multi-agent problems, and $\lambda$-return could serve as a convenient method to achieve such a trade-off.

\noindent
\textbf{Incorporating Off-Policy Data vs. Pure On-Policy Data.}
To investigate the influence of utilizing the off-policy data, we perform experiments to compare SMIX($\lambda$) against its on-policy version by scaling the size of the replay buffer. The on-policy version of SMIX($\lambda$) corresponds to SMIX($\lambda$) with buffer size $b=4$ (the most recent 4 episodes in the replay buffer are all on-policy data), while the off-policy SMIX($\lambda$) are the ones with buffer size $b>4$, where the percentage of off-policy data increases with the size of the replay buffer.

As shown in Figure \ref{fig_buffer_3s5z} and \ref{fig_buffer_2svs1sc}, all the variants of SMIX($\lambda$) incorporating off-policy data ($b>4$) perform better than the on-policy version ($b=4$) in selected scenarios.
Notably, the performance of SMIX($\lambda$) with $b=1500$ is almost twice that of the on-policy version both in terms of the final win rate and learning speed in 3s5z.
Note that 3s5z (8 units) map is more complex than 2s\_vs\_1sc (2 units) in terms of the number of agents, and consequently, the joint action space of the former is much larger.
However, more off-policy data does not always lead to better performance, as the method with $b=5000$ (green line) performs worse than the one with $b=1500$ (solid red line) in both scenarios.  Actually, the buffer size is corresponding to the $\epsilon$ in Theorem \ref{theo:matching} which measures the mismatch between the target policy $\pi$ and the behavior policy $\mu$. A smaller buffer size makes SMIX($\lambda$) less sample efficient but a larger buffer size results in a looser error bound which biases the CVF estimation. This may explain why the performance degrades once the buffer size exceeds a threshold value. And our experimental results suggest that a moderate buffer size of 1500 could be a good candidate.
\begin{figure*}[htb!]
	\centering
	\subfloat[3s5z]{\includegraphics[width=0.49\textwidth]{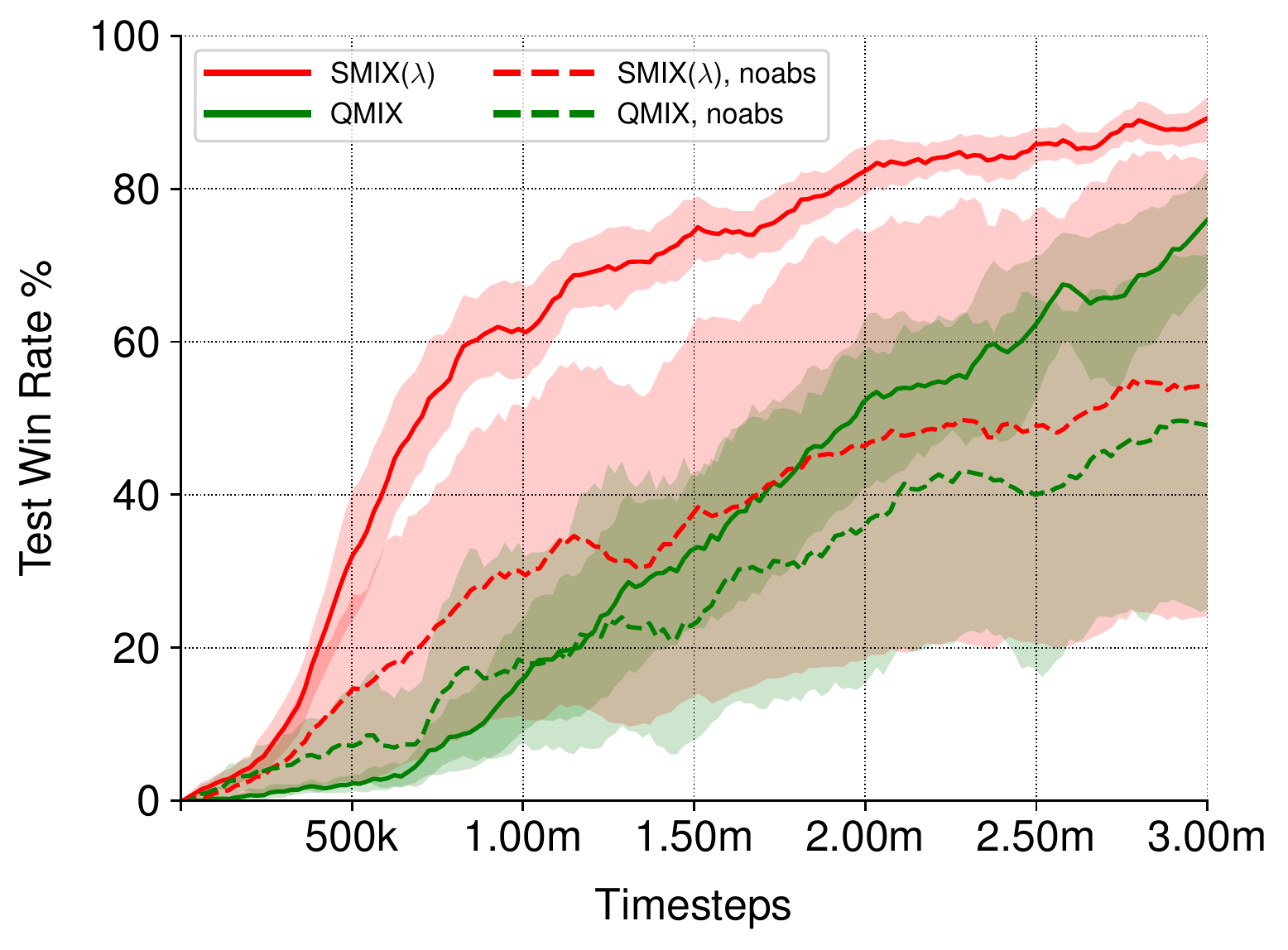}}\hfill
	\subfloat[2s\_vs\_1sc]{\includegraphics[width=0.49\textwidth]{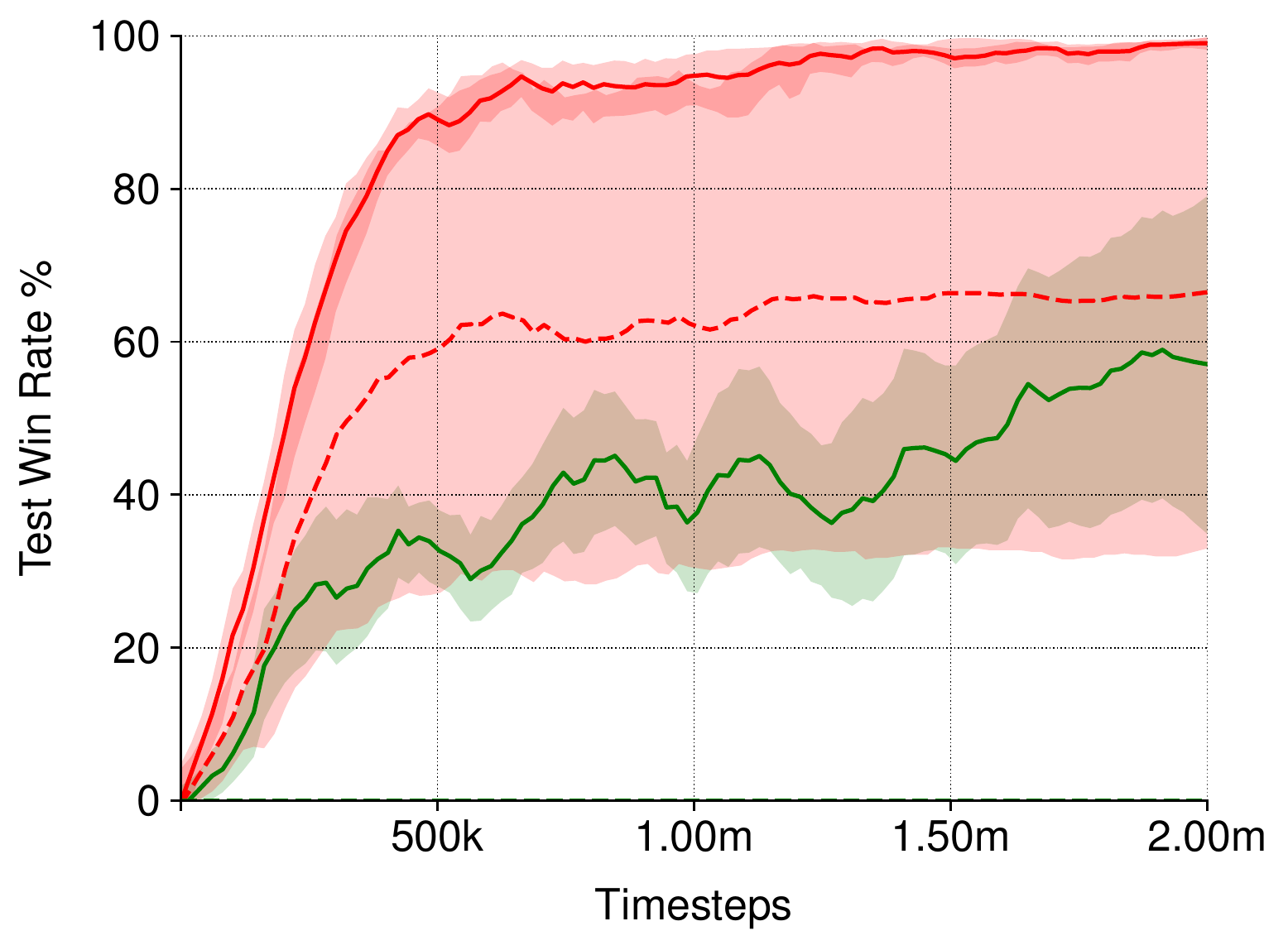}}
	\caption{{The effects of non-negative constraints of SMIX($\lambda$) and QMIX in two different scenarios. Performances of those without non-negative constraints (denoted by ``noabs") are shown as dashed lines. Methods plotted with solid lines are enforced with non-negative constraint. The mean and 95\% confidence interval is shown across 6 independent runs. }}
	\label{fig:abs}
\end{figure*}


\noindent
\textbf{Effects of Non-negative Constraints.}
{To illustrate the rationality of assumption 1, we compare the performance of SMIX($\lambda$) and QMIX with their ablations without this non-negative constraint. 

Figure \ref{fig:abs} gives comparative results of SMIX($\lambda$), and QMIX with their ablations without ``positive weights". We can see that all these algorithms suffer from drastic performance fluctuation when the additive constraint is removed. This means that non-negative constraint does play a positive role in multi-agent learning. This figure also shows that even without the requirement of non-negative weights, our SMIX($\lambda$) is still able to achieve competitive performance with the non-negative version of QMIX.}

\begin{table}[h!]
	\caption{The scalability of SMIX($\lambda$) and QMIX after training for 1 million timesteps.}
	\begin{center}
		\resizebox{0.48\textwidth}{!}{
			\begin{tabular}{|c| c||c ||c|}
				\hline
				\multicolumn{2}{|c||}{\makecell*[c]{Algorithms}}               &
				SMIX($\lambda$)                & QMIX                                                                                                                                                                                                                      \\
				\hline
				\hline
				\multirow{2}{*}{3m} & \multicolumn{1}{c||}
				{mean $\pm$ \emph{std}}& {\textbf{99}} ($\pm$0) & 95 ($\pm$3) \\
				\cline{2-4}
				&  median & \makecell[c]{\textbf{99}} & 95 \\
				
				\hline
				\multirow{2}*{8m} & \multicolumn{1}{c||}
				{mean $\pm$ \emph{std}}& \makecell[c]{\textbf{91} ($\pm$3)} & 90 ($\pm$3) \\
				\cline{2-4}
				&  median & \makecell[c]{\textbf{90}} & 89 \\
				\hline
				\multirow{2}*{25m} & \multicolumn{1}{c||}
				{mean $\pm$ \emph{std}}& \makecell[c]{\textbf{75} ($\pm$26)} & {30} ($\pm$17) \\
				\cline{2-4}
				&  median & \makecell[c]{{\textbf{93}}} & {24} \\
				\hline
			\end{tabular}
		}		
	\end{center}
	\label{table:scalability}
\end{table}
\noindent\textbf{Scalability}. The results in Table \ref{table:scalability} show the scalability of SMIX($\lambda$) and QMIX after training for 1 million steps. Overall, the performance of both methods decreases along with the increasing number of agents. However, our SMIX($\lambda$) still outperforms QMIX, especially in hard scenarios 25m. Specifically, with 3 agents (the 3m map), SMIX($\lambda$) achieves the best performance among the compared methods with a 99\% win rate. By increasing the number of agents to 8 (the 8m map), the performance of all the methods decreases due to the higher degree of challenging of the task, while our method still performs best among the compared ones\footnote{See Table \ref{table:quantitative_res} for more results on 3m and 8m.}. Finally, when the number of agents been increased to 25 (the 25m map), the performance of QMIX decreases dramatically, which is not the case for  SMIX($\lambda$). These results show that the centralized value function estimation method used in SMIX($\lambda$) method has better scalability and performs more robust in challenging tasks than QMIX. Finally, it is worth mentioning that for our experiments with up to 25 agents, the joint action space would be as large as $|\mathcal{A}|^{25}$, which imposes a great challenge to any MARL method.
\section{Conclusions \& Future work }
\label{sec:conclusions}

One of the central challenges in multi-agent reinforcement learning with CTDE settings is to estimate the centralized value function. However, the sparse experiences and {unstable} nature of the multi-agent environments make this become a challenging task. To address this issue, we present the SMIX($\lambda$) approach, by enhancing the quality of {centralized value function} in three aspects: (1) removing the greedy assumption to help to learn a more flexible functional structure, (2) using off-policy learning to alleviate the problem of sparse experiences and to improve exploration, and (3) using $\lambda$-return to balance the bias and variance of the algorithm.  Our analysis indicates that SMIX($\lambda$) has nice convergence guarantee through off-policy learning without importance sampling, which gives potential advantages in multi-agent settings. It is shown that the proposed SMIX($\lambda$) approach significantly outperforms several state-of-the-art CTDE methods on the benchmark of the StarCraft Multi-Agent Challenge, and is beneficial to other CTDE methods as well by replacing their {centralized value function} estimator with ours.

Our future work will focus on incorporating the communication and opponent modeling methods into SMIX($\lambda$) to further tackle the non-stationarity issue during the execution. We also aim to make SMIX($\lambda$) perform more efficiently in dealing with a large numbers of agents.


%

%

%
%

\ifCLASSOPTIONcaptionsoff
\newpage
\fi



\bibliographystyle{IEEEtran}
\bibliography{IEEEtran}
\end{document}